\documentclass[twocol]{ametsocV5}
\usepackage[colorlinks,citecolor=blue,linkcolor={blue},urlcolor=blue,bookmarks=false,hypertexnames=true]{hyperref} 

\usepackage{amsmath,amsfonts,amssymb,bm}
\usepackage{mathptmx}%{times}
\usepackage{newtxtext}
\usepackage{newtxmath}
\usepackage{dashrule}
\usepackage{amsmath}

\title{Overlapping boundary layers in coastal oceans}

\authors{Chao Yan, James C. McWilliams, Marcelo Chamecki \correspondingauthor{Marcelo Chamecki, chamecki@ucla.edu}}

\affiliation{Department of Atmospheric and Oceanic Sciences, University of California, Los Angeles, Los Angeles, California, USA}

\abstract{Boundary layer turbulence in coastal regions differs from that in deep ocean because of bottom interactions. In this paper, we focus on the merging of surface and bottom boundary layers in a finite-depth coastal ocean by numerically solving the wave-averaged equations using a large eddy simulation method. The ocean fluid is driven by combined effects of wind stress, surface wave, and a steady current in the presence of stable vertical stratification. The resulting flow consists of two overlapping boundary layers, i.e. surface and bottom boundary layers, separated by an interior stratification. The overlapping boundary layers evolve through three phases, i.e. a rapid deepening, an oscillatory equilibrium and a prompt merger, separated by two transitions. Before the merger, internal waves are observed in the stratified layer, and they are excited mainly by Langmuir turbulence in the surface boundary layer. These waves induce a clear modulation on the bottom-generated turbulence, facilitating the interaction between the surface and bottom boundary layers. After the merger, the Langmuir circulations originally confined to the surface layer are found to grow in size and extend down to the sea bottom (even though the surface waves do not feel the bottom), reminiscent of the well-organized Langmuir supercells. These full-depth Langmuir circulations promote the vertical mixing and enhance the bottom shear, leading to a significant enhancement of turbulence levels in the vertical column.
}

\begin{document}

%% Necessary!
\maketitle

%%%%%%%%%%%%%%%%%%%%%%%%%%%%%%%%%%%%%%%%%%%%%%%%%%%%%%%%%%%%%%%%%%%%%
% SIGNIFICANCE STATEMENT/CAPSULE SUMMARY
%%%%%%%%%%%%%%%%%%%%%%%%%%%%%%%%%%%%%%%%%%%%%%%%%%%%%%%%%%%%%%%%%%%%%
%
% If you are including an optional significance statement for a journal article or a required capsule summary for BAMS 
% (see www.ametsoc.org/ams/index.cfm/publications/authors/journal-and-bams-authors/formatting-and-manuscript-components for details), 
% please apply the necessary command as shown below:
%
% \statement
% Significance statement here.
%
% \capsule
% Capsule summary here.

%%%%%%%%%%%%%%%%%%%%%%%%%%%%%%%%%%%%%%%%%%%%%%%%%%%%%%%%%%%%%%%%%%%%%
% MAIN BODY OF PAPER
%%%%%%%%%%%%%%%%%%%%%%%%%%%%%%%%%%%%%%%%%%%%%%%%%%%%%%%%%%%%%%%%%%%%%
%

%% In all cases, if there is only one entry of this type within the higher level heading, use the star form: 
%%
\section{Introduction}\label{sec:intro}
Oceanic boundary layer flows control the turbulent mixing and mass transport in the marine environment. Depending on the forcing conditions, the turbulence therein can be classified into different regimes, i.e. (1) Langmuir turbulence in the surface boundary layer (SBL) driven by the overlying wind stress and surface gravity waves (hereafter referred to as "surface forcing") \citep{Thorpe2004ARFM, Sullivan2010ARFM, D'Asaro2014ARMS}, and (2) bottom-generated turbulence in the bottom boundary layer (BBL) owing to the drag of currents on the seafloor \citep{grant1986arfm, trowbridge2018arms}. Better understanding of oceanic turbulence and boundary layer dynamics is crucial in deriving improved parameterizations of mixing in global climate models \citep{Belcher2012grl} and regional oceanographic models \citep{Large1994RoG, McWilliams2000}.

Previous studies have mostly focused on physical processes in either the SBL of deep ocean \citep{McWiliams:1997jfm, Grant2009JPO} or the BBL over coastal regions \citep{Taylor2008b}. One of the most prominent features in the SBL is the presence of Langmuir circulations (LCs), which consist of counter-rotating vortices near the ocean surface \citep{Thorpe2004ARFM}. The interaction of wave-induced Stokes drift and wind-driven shear current give rise to these coherent structures via the Craik-Leibovich type II (CL2) instability \citep{Craik:1977JFM, Leibovich1983arfm}, which are effective at enhancing vertical mixing and turbulent exchange. The resulting Langmuir turbulence can be numerically modelled by adding a Craik-Leibovich vortex force into the momentum equation without the need to resolve the surface gravity waves \citep{Skyllingstad:1995jgr, McWiliams:1997jfm}. Compared to shear-driven turbulence, Langmuir turbulence features near-surface convergence zones with stronger turbulent fluctuations in the vertical and crosswind directions \citep{McWiliams:1997jfm, D'Asaro2001JPO, Harcourt2008JPO}.

The oceanic BBL is another dynamic part of the water column \citep{trowbridge2018arms}. In a stratified environment, the BBL structure consists of a well-mixed layer near the substrate and a stongly stable pycnocline, which is analogous to the stable atmospheric boundary layer (ABL), except that the stable ABL is always accompanied by a cooling heat flux at the ground while the seafloor is usually assumed to be adiabatic. Internal waves are generated above the pycnocline and propagate upward as a result of turbulent eddies interacting with the ambient stratification \citep{Taylor2007}. \citet{Taylor2008b} suggested that internal waves and stratification have a profound influence on the boundary layer structures. Under an oscillating tidal current, \citet{Gayen2010JFM} found out that the near-wall mixed layer grows in time with a periodic modulation by the tidal oscillation. These studies focused solely on the response of the oceanic bottom layer to the external stratification, while the dynamics associated with the ocean surface layer were not taken into account.

In shallow-water coastal regions, as the Stokes drift velocity is non-zero at the seafloor, the boundary layer turbulence differs from that in deep ocean due to the bottom interaction. The observational studies of \citet{Gargett2004Science} and \citet{Gargett2007jfm} over the inner shelf of New Jersey (water depth of 15 m) suggest that the large-scale Langmuir cells could occupy the entire water column under strong wind and wave forcing conditions. Such full-depth vortex pairs, termed Langmuir supercells (LSCs) can foster intensified near-bottom motions below the downwelling region, thereby exerting profound influences on the sediment re-suspension and mass transport \citep{Gargett2004Science}. The Large-Eddy Simulation (LES) study of \citet{Tejada-Martinez2012jfm} suggests that LSCs have the potential of interfering with the bottom log-layer dynamics. In light of this finding, \citet{Golshan:2017CF} investigated the impact of different wall treatments in LES and Reynolds-averaged Navier-Stokes (RANS) on simulation results in the presence of full-depth LCs. They suggested that the traditional wall treatment based on the log-law wall function is still valid in LES modelling. Recently, \citet{Deng:2019JFM} found out that the logarithmic layer disrupted at $Re_\tau=395$ as stated by \citet{Tejada-Martinez2012jfm} would reappear at high Reynolds number with $Re_\tau \sim \textit{O}(10^3)$, further justifying the use of log-law-based equilibrium wall models in LES studies (which lends more credibility to the use of the present wall model described in section \ref{sec:methods}).

The scenario of how Langmuir turbulence evolves becomes more intricate in the presence of enhanced bottom shear forced by mean currents associated with tides or large-scale eddies \citep{Shrestha2019JFM}. Turbulence originating near the seabed in strong tidal flows can extend to the surface over shallow well-mixed seas \citep{Smith:1999Nature}. Observations of \citet{THORPE2000} in a well-mixed water suggest that Langmuir turbulence dominates the bottom turbulence by tidal forcing when the wind speed is sufficiently larger than the current speed. \citet{Martinat2011od} quantified the flows on shallow continental shelves with a pressure gradient aligned with and perpendicular to the wind stress. With combined efforts of observations and LES, \citet{Kukulka2011jgr} found that the crosswind shear associated with the tidal currents can distort Langmuir cells in shallow water ($\sim$ 16 m). \citet{Shrestha:2018} investigated how surface forcing and downwind pressure gradient influence the length and velocity scales of LSCs in coastal zones. Recently, \citet{Shrestha2019JFM} explored how the full-depth LSCs are modulated by a range of misaligned wind-wave-current conditions. These studies have significantly advanced our understanding of Langmuir turbulence in shallow-water regions where the entire water column is turbulent.

In a sufficiently deep coastal area, the surface and bottom boundary layers are separated by an interior stable stratification, which hampers the vertical mixing across the water column. Generally, the vertical dimension of the SBL is dependent on the magnitude of surface-friction velocity and Stokes drift \citep{Grant2009JPO}, while the BBL spans a distance from the seafloor to a depth controlled by the magnitude of the current. After allowing enough time for the boundary layer development, the bottom-generated turbulence can interact with Langmuir turbulence in a nonlinear manner. For instance, the time-varying interior stratification will suppress the boundary layer growth and affect the vertical boundary layer structures accordingly \citep{Pham2017JGR, Taylor2008b}. 
Also, internal waves can be generated by the interaction of stratification with Langmuir circulations \citep{Chini2003JFM, Polton2008GRL} and turbulent motions in the BBL \citep{Taylor2007}. They play a key role in transporting energy in the ocean, regulating the boundary layer dynamics and probably driving local mixing. However, this complex flow problem is not well understood, an option we intend to address in this study. 

The major goal here is to explore how turbulence evolves in an intermediate-depth ocean where the two distinct boundary layers coexist. 
%The categorization of intermediate-depth implies that the surface waves do not feel the bottom (i.e. the wave base is above the seafloor), but the interplay between the surface and bottom boundary layers can still occur via dynamic coupling. 
Idealized LES simulations are carried out to characterize the temporal evolution of the two boundary layers and the ensuing interaction, a physical process that is crucial in determining the transport and dispersion in coastal regions \citep{grant1986arfm}. Convective turbulence driven by surface cooling and wave-induced turbulence by wave breaking increase the problem complexity, and are not considered in this study. The remaining of the paper is organized as follows. In section \ref{sec:methods}, we describe the mathematical framework, numerical techniques, and simulations set-up. The boundary layer evolution and turbulence statistics are analyzed in section \ref{sec:evolution} and \ref{sec:turb}, respecitvely. Section \ref{sec:IGW} describes the role of internal waves in transporting energy through the water column, followed by the conclusions and main findings in section \ref{sec:conclusion}.

\section{Methods}\label{sec:methods}
\subsection{Model description}
The LES technique proves to be a powerful tool in studying the boundary layer turbulence \citep{Chamecki2019RoG}. The LES framework used here solves the grid-filtered and wave-averaged equations for mass, momentum, and heat in the Boussinesq approximation (i.e. the fluid density variations are only retained in the buoyancy term). This mathematical model is first described in \citet{McWiliams:1997jfm}, which incorporates the effects of planetary rotation and advection of scalars by the Stokes drift on the basis of the original Craik-Leibovich equations \citep{Craik:1976JFM},
\begin{equation} \label{eq:continuity}
    \nabla \cdot {\widetilde{\boldsymbol{u}}}=0, 
\end{equation}
\begin{equation} \label{eq:momentum}
\begin{split}
    \frac{\partial{\widetilde{\boldsymbol{u}}}}{\partial{t}}+{\widetilde{\boldsymbol{u}}} \cdot \nabla{\widetilde{\boldsymbol{u}}} =&-\nabla{\Pi}-f\boldsymbol{e}_z\times\left(\widetilde{\boldsymbol{u}}+\boldsymbol{u}_s-\boldsymbol{u}_g\right) \\
    &+\boldsymbol{u}_s \times \widetilde{\boldsymbol{\zeta}}+
    \left(1-\frac{\widetilde{\rho}}{\rho_0}\right)g\boldsymbol{e}_z + 
    \nabla \cdot {{\boldsymbol{\tau}}^d},
\end{split}
\end{equation}
\begin{equation} \label{eq:temperature}
    \frac{\partial{\widetilde{\theta}}}{\partial{t}}+ \left(\widetilde{\boldsymbol{u}}+\boldsymbol{u}_s\right) \cdot \nabla \widetilde{\theta}=\nabla \cdot {\boldsymbol{\tau}_\theta},
\end{equation}
Here, the tilde indicates the grid-filtered variables, $\widetilde{\rho}$ is the density of seawater, $\rho_0$ is the reference density, $\widetilde{\theta}$ is the potential temperature, $g$ is the acceleration of gravity. The changes in the density $\widetilde{\rho}$ is assumed to be caused by $\widetilde{\theta}$ via an inverse relationship, i.e. $\rho=\rho_0[1-\alpha(\theta-\theta_0)]$, where $\alpha=2\times10^{-4}\ \mathrm{K}^{-1}$ is the thermal expansion coefficient, and $\theta_0$ is the reference temperature. In a Cartesian coordinate system $\boldsymbol{x}=({x}, {y}, {z})$, $\boldsymbol{e}_z$ is the unit vector in the vertical direction, and the filtered velocity vector ${\widetilde{\boldsymbol{u}}}=({\widetilde{u}, \widetilde{v}, \widetilde{w}})$ denotes the velocity components in the streamwise ($x$), crosswise ($y$), and vertical ($z$) directions. The vertical coordinate is defined positive upward with $z=0$ at the ocean surface.

In equation \eqref{eq:momentum}, $f$ is the Coriolis frequency, $\boldsymbol{u}_{s}$ is the Stokes drift associated with surface waves, and a geostrophic current $\boldsymbol{u}_{g}$ is generated by imposing a pressure gradient $f\boldsymbol{u}_{g}$ to represent the effect of mesoscale eddies. Though a non-rotational LES of LSCs yields good agreement with observations in shallow coastal ocean \citep{Tejada-Martinez:2007jfm, Grosch2016jpo}, we still include the Coriolis term in our simulations to better represent the real ocean flow.  
The viscosity is assumed to be negligible for high-Reynolds number flows considered in the present study. The third term on the right-hand-side (RHS) of \eqref{eq:momentum} is the Craik-Leibovich vortex force $\boldsymbol{u}_s \times \widetilde{\boldsymbol{\zeta}}$, where $\widetilde{\boldsymbol{\zeta}}=\nabla \times \widetilde{\boldsymbol{u}}$ is the vorticity. ${\boldsymbol{\tau}^d}$ is the deviatoric part of the subgrid-scale (SGS) stress tensor $\boldsymbol{\tau} (=\widetilde{\boldsymbol{u}}\widetilde{\boldsymbol{u}}-\widetilde{\boldsymbol{u}\boldsymbol{u}})$, i.e. ${\boldsymbol{\tau}^d}={\boldsymbol{\tau}}-\frac{1}{3}\mathrm{tr}(\boldsymbol{\tau})\ \boldsymbol{I}$, with $\mathrm{tr}(\boldsymbol{\tau})$ being the trace of $\boldsymbol{\tau}$, and $\boldsymbol{I}$ is the identity tensor. $\Pi=\widetilde{p}/{\rho_0}+\frac{1}{3}\mathrm{tr}({\boldsymbol{\tau}})+\frac{1}{2}\left| {\widetilde{\boldsymbol{u}} + \boldsymbol{u}_{s}} \right|^2 - \frac{1}{2}\left|\widetilde{\boldsymbol{u}}\right|^2$ is the generalized pressure, with $\widetilde{p}$ being the resolved pressure.

The SGS stress tensor $\boldsymbol{\tau}$, together with the SGS thermal flux vector $\boldsymbol{\tau}_\theta=\widetilde{\boldsymbol{u}}\widetilde{\theta}-\widetilde{\boldsymbol{u}\theta}$ in \eqref{eq:temperature}, account for the effect of unresolved turbulence, and they are modeled using Smagorinsky's eddy viscosity model, i.e.
\begin{equation} \label{eq:SGS model}
    \tau_{ij}=2\nu_t\widetilde{S_{ij}}, \quad
    \tau_{\theta j}=\frac{\nu_t}{Pr_t}\frac{\partial{\widetilde{\theta}}}{\partial{x_j}}, \quad
    \nu_t=(C_s\Delta)^2\sqrt{2\widetilde{S_{ij}}\widetilde{S_{ij}}}.
\end{equation}
Here, $\nu_t$ is the SGS eddy viscosity, $Pr_t$ is the turbulent Prandtl number, $\Delta$ is the grid filter size, $\widetilde{S_{ij}}=({\partial{\widetilde{u_i}}}/{\partial{x_j}}$ + ${\partial{\widetilde{u_j}}}/{\partial{x_i}})/2$ is the resolved strain-rate tensor, $C_s$ is the subgrid model coefficient determined using the Lagrangian scale-dependent dynamic Smagorinsky SGS model \citep{Bou-Zeid:2005pof}. The SGS heat flux $\boldsymbol{\tau}_\theta$ is then parameterized using an eddy diffusivity closure as shown in \eqref{eq:SGS model} with a prescribed value of $Pr_t=0.4$.

The surface wave motions are not explicitly resolved in our simulations, instead, the Stokes drift velocity $\boldsymbol{u}_{s}$ is added to the governing equations to reflect the effect of orbital motions of surface waves upon the mean currents. Here, we only consider a steady monochromatic wave representative of the wave field observed in nature. Assuming the surface gravity wave propagates along the mean wind direction (i.e. $x$ direction), the Stoke drift velocity has a form $\boldsymbol{u}_{s}=(u_s(z), 0, 0)$, where $u_s$ is given by \citep{Tejada-Martinez:2007jfm},
\begin{equation} \label{eq:stokes drift}
    u_{s}=U_s\frac{\textrm{cosh}\left[{2k(z+H)}\right]}{2\textrm{sinh}^2(kH)},
\end{equation}
in which $k$ is the wavenumber, $U_s=\sigma_w k a_w^2$ is the characteristic value of $u_s$ with $\sigma_w$ and $a_w$ being the frequency and wave amplitude respectively.

Periodic boundary conditions are imposed in the horizontal directions assuming that the flow is horizontally homogeneous and free from the coastline complexities. This periodicity assumption is valid for small coastal regions with flat bottom slope and uniform forcing conditions \citep{Burchard2008PiG}. The top boundary is specified as a non-deforming frictionless surface subject to a constant wind shear stress and zero buoyancy flux. To avoid the need to resolve the near-wall turbulent motions (as the associated turbulent length scale is very small), an equilibrium wall model based on the log-law (valid at high $Re$ number according to \citet{Deng:2019JFM}) is adopted to calculate the wall-friction stresses $\tau_{b,i} (i=1, 2)$ using the resolved velocity field at the first grid level $z_p = {\Delta}z/2$ (${\Delta}z$ is the vertical resolution) above the wall, i.e.
\begin{equation} \label{eq:wall model}
    \frac{\tau_{b,i}}{\rho}=-{u_{*b}^2}=-\left[\frac{\kappa{U}}{\ln(z_p/z_0)}\right]^2\frac{\Breve{\widetilde{u_i}}}{U},\qquad i = 1,\ 2,
\end{equation}
This wall model involves an additional test filtering operation (denoted by a breve $\Breve{\cdot}$) described in \citet{Bou-Zeid:2005pof}, and $U$ is the magnitude of the local test-filtered velocity, $\kappa = 0.4$ is the Von Karman constant, $u_{*b}$ is the friction velocity at the bottom wall, $z_0$ is the bottom roughness length that may influence the bed friction and affect the development of BBL turbulence. Here, we assume that the seafloor is adiabatic and has a roughness length of $z_0 = 0.01$ m, a typical value for areas of sandy substrate \citep[see supplementary data in][]{Jones:2015em}.

Spatial derivatives in the horizontal directions are treated with pseudo-spectral differentiation, while the derivatives in the vertical direction are discretized using a second-order central-difference scheme. The aliasing errors associated with the non-linear terms are removed based on the 3/2 rule. Time advancement is performed using the fully explicit second-order accurate Adams-Bashforth scheme. The numerical code has been validated against simulations of Langmuir turbulence in deep ocean \citep{McWiliams:1997jfm}, and applied to modeling developing boundary layer flow over a marine macroalgal farm \citep{Yan:2020}. For simplicity, the tilde symbols used to denote resolved variables are omitted hereafter.

\subsection{Simulation set-up} \label{sec:simulation}

\begin{figure}[t]
    \centerline{\includegraphics[width=1.0\linewidth]{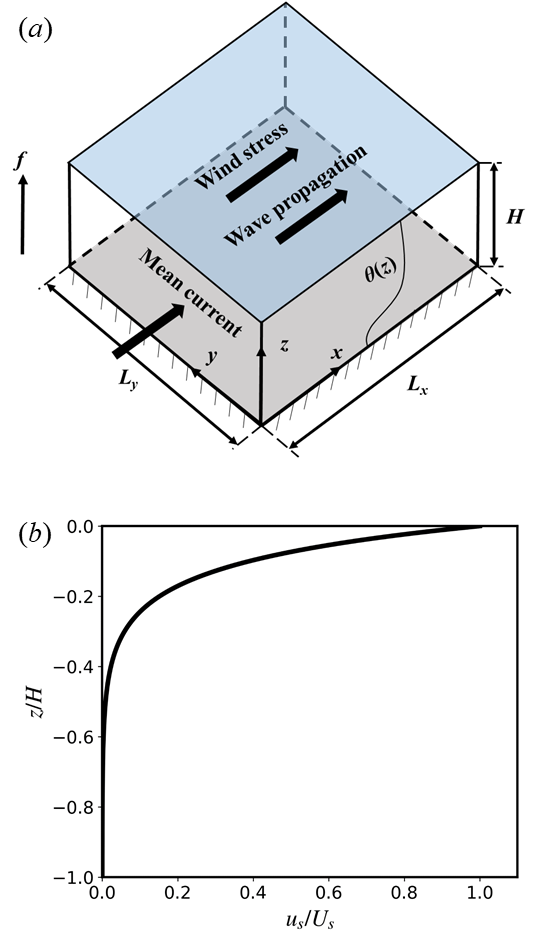}}
    \caption{(\textit{a}) Schematic of the computational model for turbulent flow in shallow water under the effects of wind, wave, current and stratification; (\textit{b}) Distribution of the wave-induced Stokes drift velocity $u_s/U_s$ in the vertical column.}
\label{fig:domain}
\end{figure}

The flow is driven by two main forcings, i.e. a surface forcing and a geostrophic current, see figure \ref{fig:domain}\textit{a}. A constant wind stress $\tau_s=0.148\ \mathrm{N\ m^{-2}}$ is applied at the air-sea surface and is aligned with the streamwise $x$-direction. The corresponding wind speed at 10-m height is $U_{10}=10\ \mathrm{m\ s^{-1}}$, and the friction velocity at the ocean surface is $u_{*s}=1.22\times10^{-2}\ \mathrm{m\ s^{-1}}$. The monochromatic surface wave is propagating along the $x$-direction, with a wavelength of $\lambda=60\ \mathrm{m}$ and an amplitude of $a=0.8\ \mathrm{m}$, yielding $U_s=0.136\ \mathrm{m\ s^{-1}}$ and $La_t=0.3$. 
These parameter values represent typical wind and wave conditions in coastal regions \citep{Belcher2012grl}. The geostrophic current $\boldsymbol{u}_{g}=(u_g, 0, 0)$ is aligned in the $x-$direction and remains constant over time, assuming that the variations of mesoscale flow features and tidal forcing are negligible on the time scale of interest here. For comparison, the flows driven by either the surface forcing or the geostrophic current are also simulated. %to assess how bottom-generated turbulence in BBL interacts with Langmuir turbulence in SBL.

The water depth is $H=45$m, and thus the Stokes drift velocity \eqref{eq:stokes drift} is identically zero in the lower half of the water column (see figure \ref{fig:domain}\textit{b}) rather than persisting towards the bottom wall as in shallow-water Langmuir turbulence \citep{Gargett2004Science, Tejada-Martinez:2007jfm}. It is worth mentioning that observations at a site off Georgia (27-m-deep) suggested that the surface layer LCs will not evolve into full-depth LSCs when the water depth is much deeper than 25-30 m \citep{Gargett2014jmr}. The computational domain size in the horizontal direction is $L_x=L_y=2\pi H$, which is assumed to be large enough to minimize the influence of the finite domain size \citep{Shrestha:2018}. The mesh is uniformly distributed in all three directions, and the computational parameters and grid resolution are shown in table \ref{tab:1}. All the simulations start as uniformly stratified fluid (USF), i.e. temperature is linearly stratified throughout the entire water column with a initial temperature gradient ${\mathrm{d}\theta/\mathrm{d}z}\vert_0 = 0.1\ \mathrm{K\ m^{-1}}$. The mean velocity $U=u+iv$ is initialized with the steady-state bottom Ekman layer solution \citep{Wyngaard:2010},
\begin{equation} \label{eq:init}
    U = u_g\left(1-e^{-\beta z}\mathrm{cos}{\beta z}\right) + i u_g e^{-\beta z}\mathrm{sin}{\beta z}
\end{equation}
in which $\beta=(f/2\nu_e)^{1/2}$, and $\nu_e=10^{-4}\ \mathrm{m^2\ s^{-1}}$ is the effective eddy viscosity in the bottom Ekman spiral. The subscript in table \ref{tab:1} indicates different flow regimes, i.e. $\left(\cdot\right)_\mathrm{S\&B}$ denotes the co-existence of SBL and BBL, while $\left(\cdot\right)_\mathrm{SBL}$ or $\left(\cdot\right)_\mathrm{BBL}$ implies the simulation in which only SBL or BBL is present. The simulations are carried out for $t/T_f=12$ time units, where $T_f=2\pi/f$ is the inertial period, with a dimensional time step $t=0.15\ \mathrm{s}$ (i.e. the integration time is more than 5 million time steps). To capture the boundary layer evolution, the flow and thermal fields are decomposed into a horizontal mean (denoted with angle brackets) and deviations from it (denoted with a single prime), e.g. $\boldsymbol{u} = \langle \boldsymbol{u} \rangle + \boldsymbol{u}^{\prime}$. When the flow reaches a quasi-equilibrium state, an additional time-averaging operation, denoted by an overbar (e.g. $\overline{\langle \boldsymbol{u} \rangle}$), is taken over an inertial period $T_f$ so as to mitigate the effect of inertial oscillations.

\begin{table*}[t]
\caption{Relevant parameters of the LES runs.}\label{tab:1}
\begin{center}
\begin{tabular}{lcccccccc}
\hline\hline
Case  & {$u_{*s}\ (\mathrm{cm\ s^{-1}})$} & $\lambda\ (\mathrm{m})$ & $a\ (\mathrm{m})$ & $La_{t}$ & {${u_{g}\ (\mathrm{m\ s^{-1}})}$} &{${\mathrm{d}\theta /\mathrm{d}z \vert_0}\ (\mathrm{K\ m^{-1}})$} & $L_x \times L_y \times L_z$ & $N_x \times N_y \times N_z$\\
$\mathrm{USF_{S\&B}}$ & 1.22 & 60 & 0.8 & 0.3 & 0.25 & 0.1 & $2\pi H \times 2\pi H \times H$ & $256 \times 256 \times 144$ \\
$\mathrm{USF_{SBL}}$ & 1.22 & 60 & 0.8 & 0.3 & 0 & 0.1 & $2\pi H \times 2\pi H \times H$ & 256 $\times$ 256 $\times$ 144 \\
$\mathrm{USF_{BBL}}$ & {0} & {N/A} & 0 & {N/A} & 0.25 & 0.1 & $2\pi H \times 2\pi H \times H$ & 256 $\times$ 256 $\times$ 144 \\
\hline
\end{tabular}
\end{center}
\end{table*}

\begin{figure*}[t]
    \centerline{\includegraphics[width=0.8\linewidth]{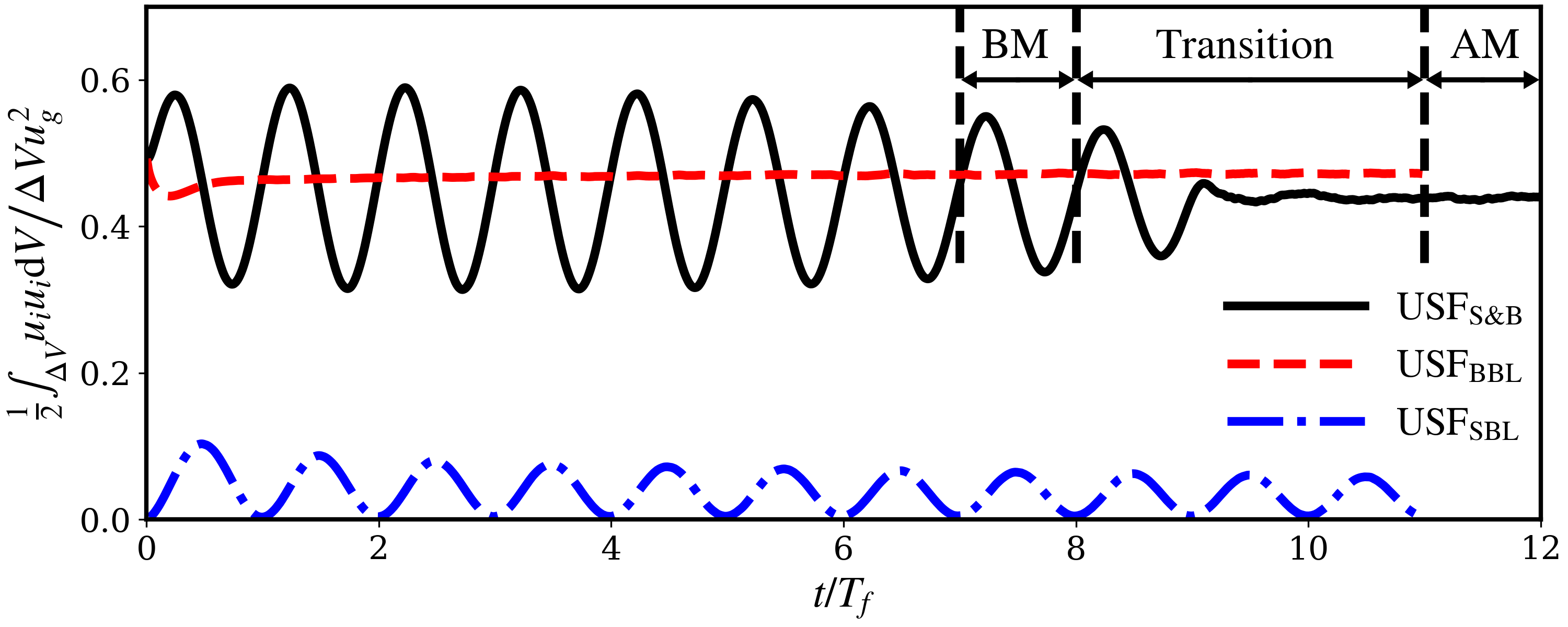}}
    \caption{Variations of the volume-averaged kinetic energy over time (in terms of the inertial period $T_f$) for all the three simulations in table \ref{tab:1}. Here, $u_g = 0.25\ \mathrm{m\ s^{-1}}$ is used as the scaling velocity, even for $\mathrm{USF_{SBL}}$ where there is no mean current to drive the flow. The surface and bottom boundary layers for $\mathrm{USF_{S\&B}}$ will eventually merge, transitioning from a quasi-steady status to a different equilibrium state. The LES solutions averaged over two separate inertial periods, one before the merger (BM) and one after (AM), are assumed to be representative of the pre-merger and post-merger regimes, respectively.}
\label{fig:ke}
\end{figure*}

Figure \ref{fig:ke} shows the times series of the volume-averaged kinetic energy from the three LES simulations included in this study, i.e. $\mathrm{USF_{S\&B}}$ (black solid line), $\mathrm{USF_{SBL}}$ (blue dash-dotted line), and $\mathrm{USF_{BBL}}$ (red dashed line). Here, we use the same normalization factor $u_g = 0.25\ \mathrm{m\ s^{-1}}$ for all three cases even though there is no geostrophic current in $\mathrm{USF_{SBL}}$ to drive the flow. In a rotating environment, slowly decaying inertial oscillations of the horizontal current are observed for $\mathrm{USF_{S\&B}}$ and $\mathrm{USF_{SBL}}$. The kinetic energy in $\mathrm{USF_{SBL}}$ goes to nearly zero at the end of each inertial period because this simulation actually starts from rest (equation \eqref{eq:init}). For case $\mathrm{USF_{S\&B}}$, the two boundary layers will eventually merge as the surface and bottom mixed layers both grow continually into the interior stratified layer. Accordingly, the flow regime transitions from a quasi-steady status (for $t/T_f<8$) to a different equilibrium state (for $t/T_f>11$). Test runs suggest that the transition period will be delayed or accelerated depending on the magnitude of the surface forcing, geostrophic current, and bottom roughness length, but the flow sensitivity to these parameters is not pursued here.

Interestingly, case $\mathrm{USF_{BBL}}$ and the final equilibrium regime of $\mathrm{USF_{S\&B}}$ are seemingly exempt from any inertial oscillations (figure \ref{fig:ke}). This is because, when the entire boundary layer is influenced by the bottom floor, the wall friction causes damping and modifies the restoring force (i.e. Coriolis force) of inertial oscillations \citep{Schroter2013QJRMS}, thus the amplitudes of inertial motions are very small, similar to that in the ABL \citep{Lundquist2003JAS}. The LES solutions for $\mathrm{USF_{S\&B}}$ are averaged over two separated inertial periods, denoted as BM and AM in figure \ref{fig:ke}, to examine the variation of turbulent dynamics before and after the merger between the SBL and BBL. Statistics from $\mathrm{USF_{SBL}}$ and $\mathrm{USF_{BBL}}$, averaged over BM, are also extracted for comparison.

\begin{figure}[t]
    \centerline{\includegraphics[width=0.8\linewidth]{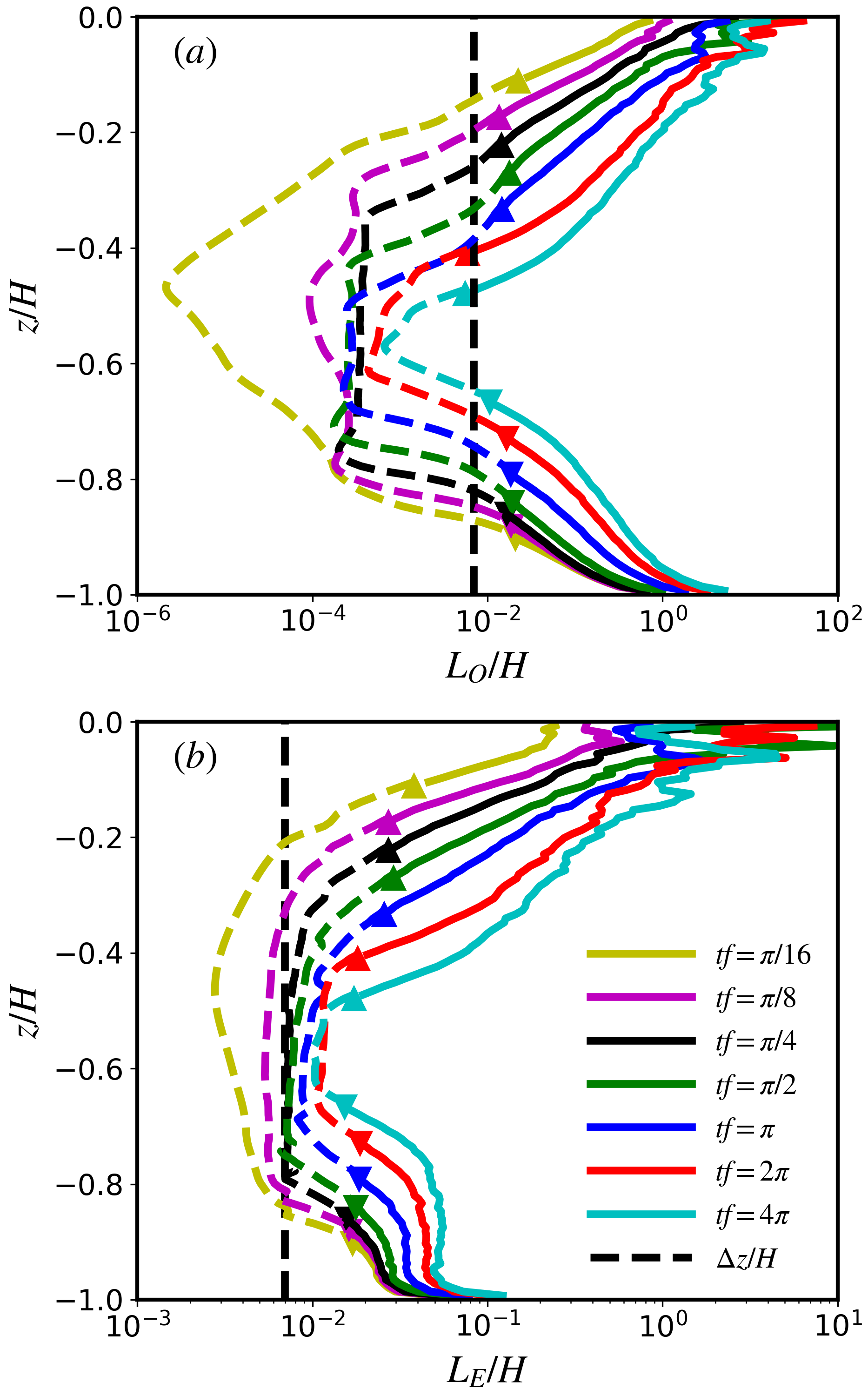}}
    \caption{The vertical profiles of (\textit{a}) the Ozmidov scale $L_O$, and (\textit{b}) the Ellison scale $L_E$ within the first two inertial periods for case $\mathrm{USF_{S\&B}}$. The up- and down-pointing triangles denote the boundary layer depths at each time moment for SBL and BBL, respectively. The section between the two boundary layers are plotted with a dashed line. The vertical dashed line marks the vertical grid resolution. }
    \label{fig:reso}
\end{figure}

Here, we define the depth of upper and lower boundary layers to be, respectively, the vertical levels where the potential temperature exceeds a certain percentage of the temperature in the upper and lower mixed layers \citep[adapted from the temperature contour method in][]{Sullivan:1998jas}, 
\begin{equation} \label{eq:mld}
  z_i = \{z: \left| \langle \theta \rangle (z) - \theta_\mathrm{ML} \right| = \chi \theta_\mathrm{ML}\}
\end{equation}
where $\chi$ is a predefined constant. In general, our resolution is reasonable to resolve internal waves, but not fine enough to capture wave breaking. To confirm that the grid resolution is sufficient to resolve the key flow features in the boundary layers, we compare the vertical grid spacing and two relevant length scales given below, i.e. the Ozmidov scale $L_O$ and the Ellison scale $L_E$, 
\begin{equation} \label{eq:reso}
    L_O = \varepsilon^{1/2}/N^{3/2}, \qquad 
    L_E = \frac{{\langle \theta'^{2} \rangle}^{1/2}}{{\mathrm{d}\langle \theta \rangle/\mathrm{d}z}}
\end{equation}
in which $\varepsilon$ is the rate of turbulent kinetic energy (TKE) dissipation (estimated from the SGS dissipation as the viscosity is omitted here), and $N=\sqrt{\alpha g \cdot {\mathrm{d}\langle\theta\rangle/\mathrm{d}z}}$ is the buoyancy frequency. The Ozmidov scale $L_O$ gives an estimate of the smallest scale of turbulent eddies influenced by stratification \citep{Smyth:2000PoF}, while the Ellison scale $L_E$ represents the scale of boundary layer eddies responsible for entrainment \citep{Taylor2008b}. Figure \ref{fig:reso} shows the vertical profiles of $L_O$ and $L_E$ within the first two inertial periods for case $\mathrm{USF_{S\&B}}$. Outside of the boundary layers, $L_O$ becomes irrelevant because the flow is mostly non-turbulent (even in the presence of internal waves), and %strongly influenced by stratification. Now that the flow in the stratified layer is viscous-dominated, the use of SGS dissipation in calculating $L_O$ is then inappropriate and may cause inaccuracies. Thus, 
the section of $L_O$ profile within the stratified layer is highlighted by dash-dotted lines. As the flow evolves, we can see that the present vertical grid resolution (black dashed line) is sufficient to resolve the local Ozmidov and Ellison scales near the outer edges of the boundary layers (up- and down-pointing triangles). This suggests that the present simulations are able to capture the boundary layer growth due to entrainment, thus lending confidence to the accuracy of the LES solutions.

\begin{figure*}
    \centerline{\includegraphics[width=1.0\linewidth]{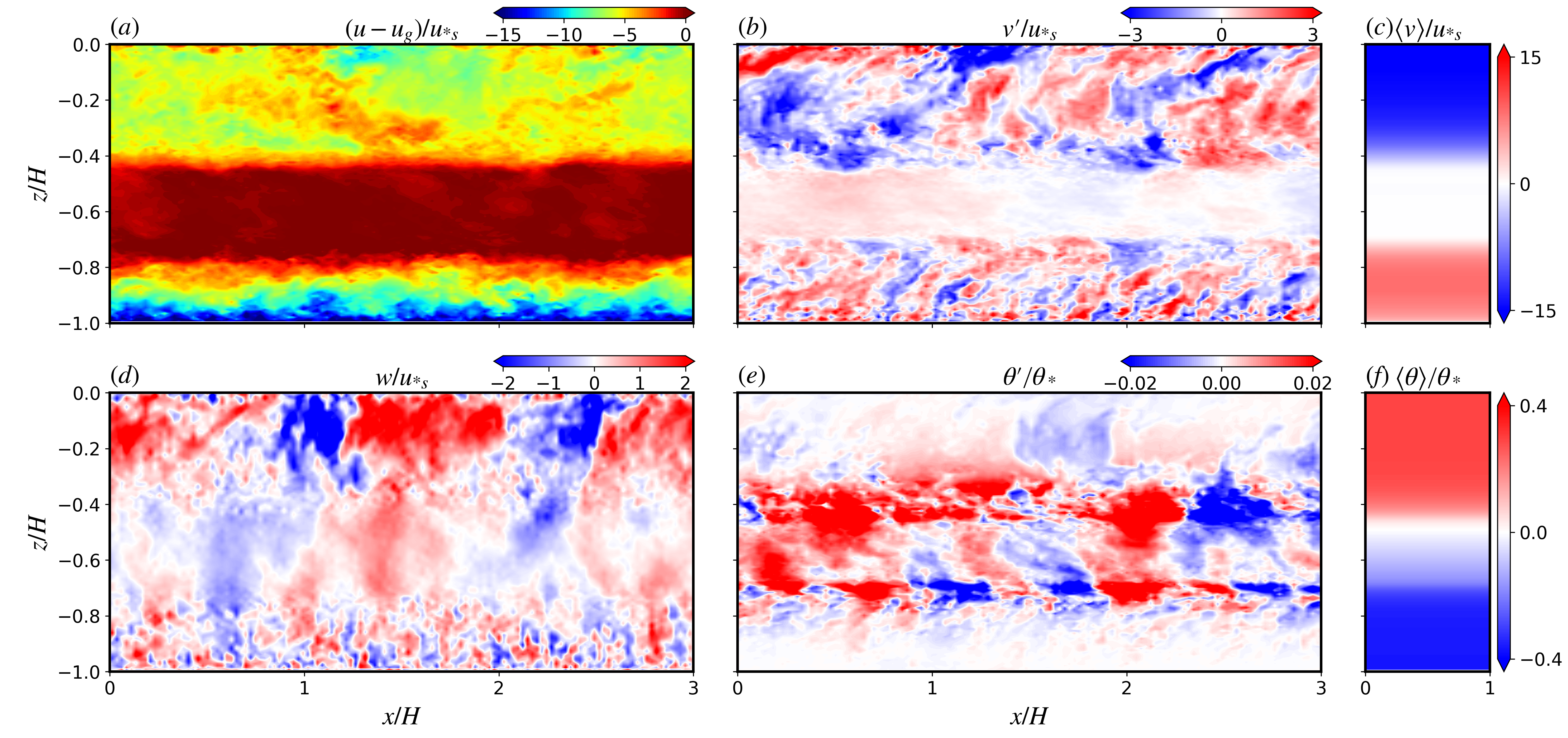}}
    \caption{Snapshot of the (\textit{a}) ageostrophic streamwise velocity $(u-u_g)/u_{*s}$, (\textit{b}) {\color{red}fluctuating component of crosswise velocity $v^{\prime}/u_{*s}$} and (\textit{c}) horizontal-averaged crosswise velocity $\langle v \rangle/u_{*s}$, (\textit{d}) vertical velocity $w/u_{*s}$, and (\textit{e}) {\color{red}fluctuating component of potential temperature $\theta/\theta_*$} and (\textit{e}) horizontal-averaged potential temperature $\langle \theta \rangle/\theta_*$ in the longitudinal $x-z$ plane at $t/T_f=0.5$ for case $\mathrm{USF_{S\&B}}$. Only a fraction of the horizontal domain is shown here.}
    \label{fig:snapshot}
\end{figure*}

\begin{figure*}
    \centerline{\includegraphics[width=1.0\linewidth]{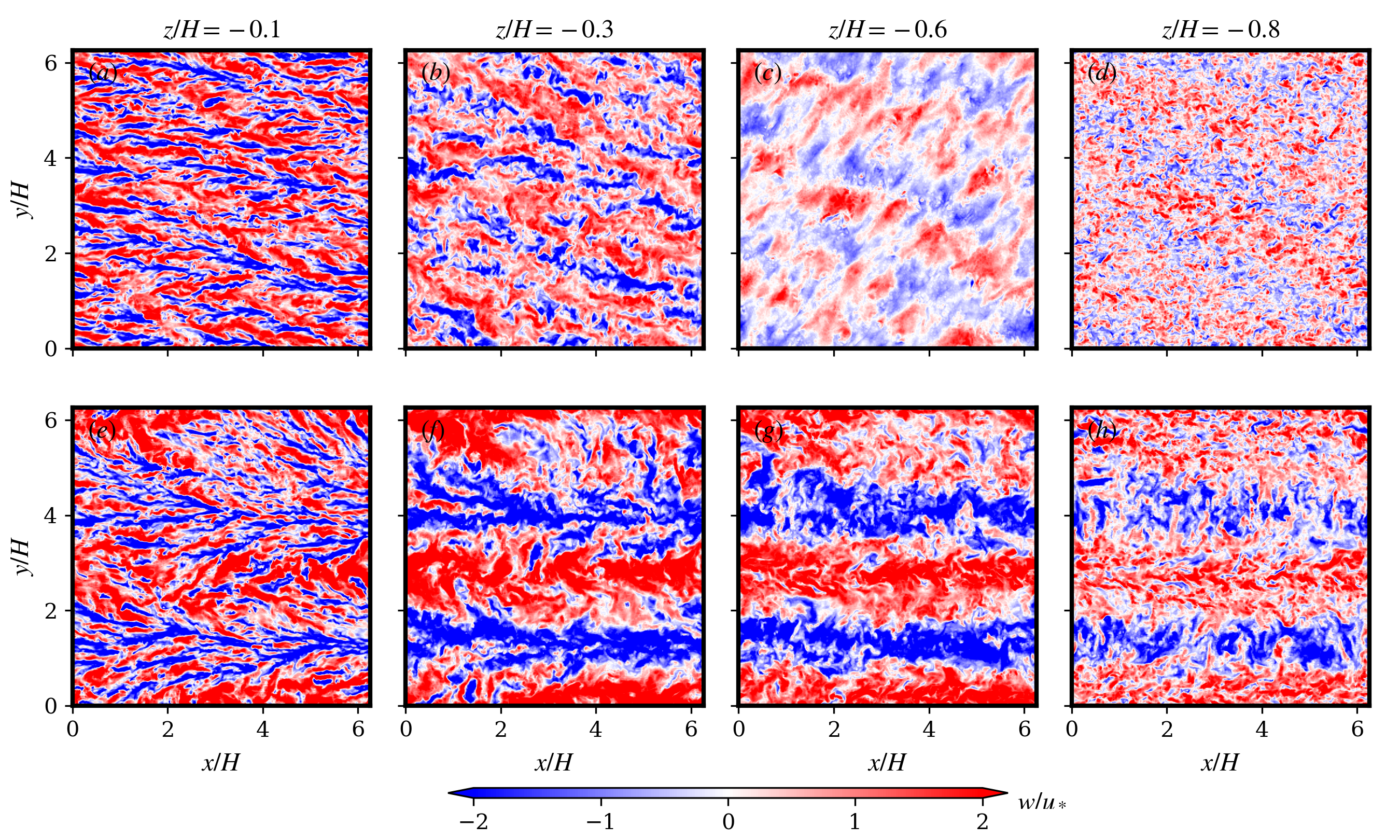}}
    \caption{Snapshots of the vertical velocity $w/u_{*s}$ at four different depths, i.e. $z/H=-0.1,\ -0.3,\ -0.6,\ -0.8$, at $t/T_f=4.5$ for case $\mathrm{USF_{S\&B}}$: (upper panels) $t/T_f=7.88$ (i.e. before the merger) and (lower panels) $t/T_f=10$ (i.e. after the merger).}
    \label{fig:snapshot_xy}
\end{figure*}

% ---
\section{Temporal evolution of the boundary layers} \label{sec:evolution}
\subsection{Visualization of the overlapping boundary layers} \label{sec:visual}

\begin{figure*}
    \centerline{\includegraphics[width=0.8\linewidth]{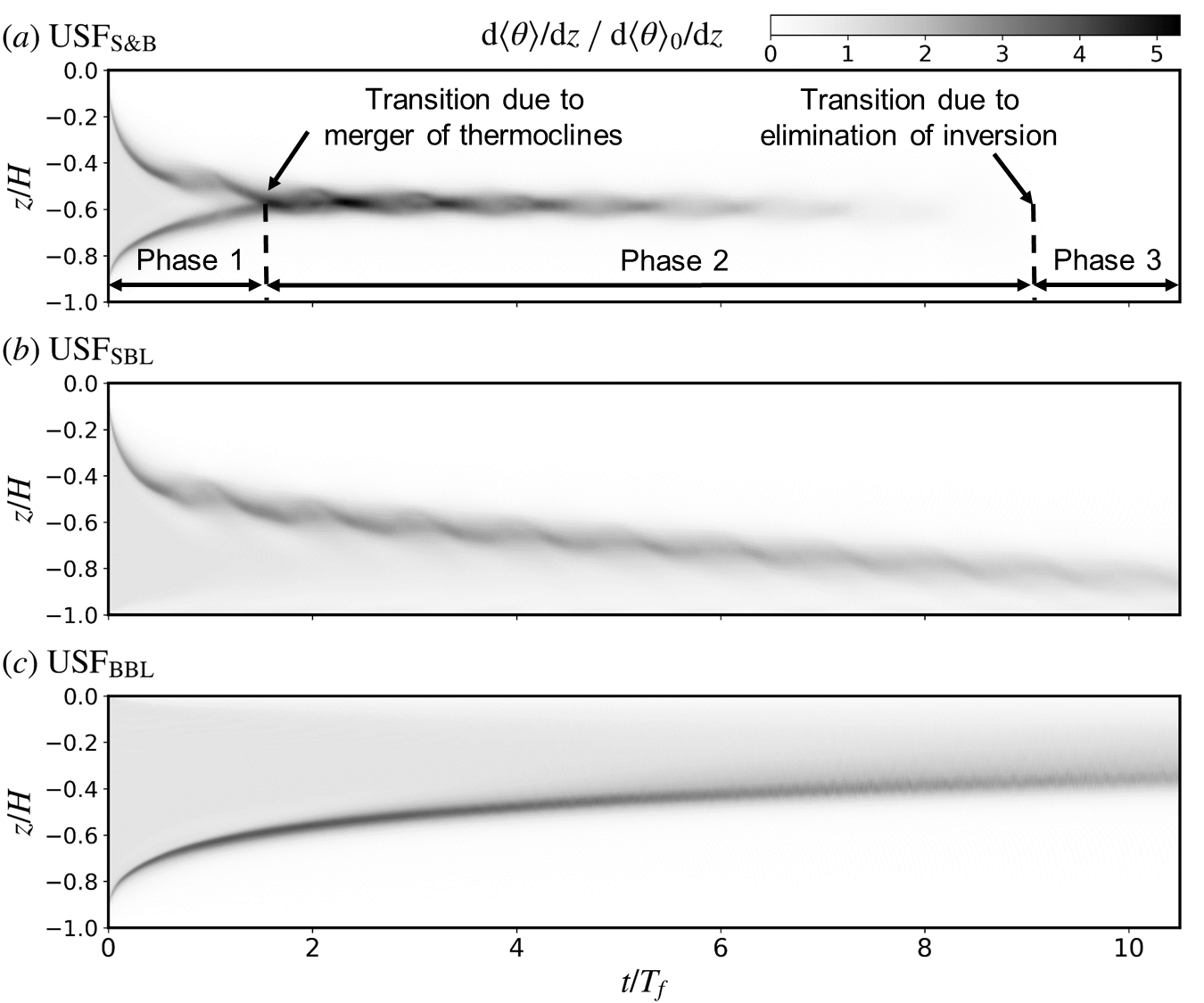}}
    \caption{Evolution of the plane-averaged temperature gradient ${\mathrm{d}\langle \theta \rangle/\mathrm{d}z}$, normalized by its initial value ${\mathrm{d}\langle \theta \rangle_0/\mathrm{d}z}$, for the three scenarios of coastal boundary layer flow: (\textit{a}) $\mathrm{USF_{S\&B}}$; (\textit{b}) $\mathrm{USF_{SBL}}$; (\textit{c}) $\mathrm{USF_{BBL}}$.}
\label{fig:dtdz}
\end{figure*}

Figure \ref{fig:snapshot} shows the instantaneous $x-z$ slices of the velocity components (normalized by $u_{*s}$) and potential temperature (normalized by $\theta_*=H{\mathrm{d}\theta /\mathrm{d}z \vert_0}$) at $t/T_f=0.5$ for simulation $\mathrm{USF_{S\&B}}$. To better visualize the turbulent fluctuations, the instantaneous fields of $v$ and $\theta$ are both decomposed into a fluctuating component and a horizontal averaged component as noted in the caption. Only the temperature deviation from the bulk temperature is considered here (figure \ref{fig:snapshot}\textit{f}). In figure \ref{fig:snapshot}, we can clearly observe a three-layer structure in the vertical column. The turbulent eddy motions, such as Langmuir circulations and bottom-generated turbulence, are mainly confined to the upper and lower boundary layer (i.e. SBL and BBL), while the central stratified layer is mostly non-turbulent and stays approximately in geostrophic balance (figure \ref{fig:snapshot}\textit{a} and \textit{b}). The streamwise velocity $u$ (figure \ref{fig:snapshot}\textit{a}) is reduced in the SBL because the current driven by the surface forcing is opposed to the Stokes drift, a typical feature of Langmuir turbulence in the upper surface layer \citep{McWiliams:1997jfm}. The crosswise velocity $v$ in the SBL and BBL (figure \ref{fig:snapshot}\textit{c}) is directed to the right of the wind stress (positive $x-$direction) and bottom stress (negative $x-$directions), respectively, which is consistent with the surface and bottom Ekman spirals in the oceanic boundary layer flow \citep{Taylor2007, Pham2017JGR}. The alternating downward and upward $w$ in the stratified layer in figure \ref{fig:snapshot}\textit{d} implies the propagation of internal waves. As the forcing conditions are kept constant and there is no topography, these internal waves are believed to be excited by boundary layer turbulence. Internal waves can potentially alter the dynamics and energetics of boundary layers by transporting momentum and energy in the vertical \citep{Chini2003JFM, Taylor2007}, thus facilitating dynamical coupling between the upper and lower boundary layers. The boundary layer turbulence continually erodes the stratification, and homogenizes the temperature field in the surface and bottom waters (figure \ref{fig:snapshot}\textit{e}). Since no buoyancy flux across the air-sea interface is present to stabilize the temperature profile, the thermal field will keep evolving over time till the temperature is eventually well mixed throughout the entire water column.

Figure \ref{fig:snapshot_xy} shows the instantaneous field of $w/u_*$ in the $x-y$ planes at four different vertical levels ($z/H=-0.1,\ -0.3,\ -0.6,\ -0.8$) at $t/T_f=7.88$ (before the merger, upper panels) and $t/T_f=10.0$ (after merger, lower panels). Before the two boundary layers merge, the elongated streaks of downwelling velocity (colored by blue) observed in the upper layers (figure \ref{fig:snapshot_xy}\textit{a} and \textit{b}) are signatures of Langmuir turbulence similar to those found in the deep ocean, where Langmuir circulations are oriented to the right of the wind direction \citep{McWiliams:1997jfm}. In the mid-layer (figure \ref{fig:snapshot_xy}\textit{c}), we can observe quasi-periodic propagating variations of the vertical velocity, indicating the presence of internal waves. The flow field in figure \ref{fig:snapshot_xy}\textit{d} displays evident spatial correlation with that in figure \ref{fig:snapshot_xy}\textit{c}, which suggests that the internal waves impose their imprint on the boundary layer turbulence near the bottom wall. However, this wave pattern in the stratified layer does not persist over time but it is characterized by intermittent behaviors, i.e. the internal waves constantly disappear and reappear with varying direction of propagation (see the supplementary movie). This is because internal waves of different frequencies and wave amplitudes interfere with each other, thus occasionally smearing out any persistent wave patterns. After the two boundary layers merge, a strongly coherent pattern with upwelling and downwelling velocity alternating periodically in the crosswise direction is clearly seen throughout the water column (figure \ref{fig:snapshot_xy}\textit{e}-\textit{h}), indicating the presence of two large-scale counter-rotating vortex pairs that are reminiscent of the full-depth LSCs \citep{Gargett2004Science, Tejada-Martinez:2007jfm}. From figure \ref{fig:snapshot_xy}\textit{h}, the full-depth Langmuir circulations clearly modulate the BBL dynamics \citep{Deng:2019JFM, Shrestha2019EFM}. As the simulation evolves further in time, these two downwelling regions merge into one and the whole domain fits only one pair of counter-rotating vortices (more evidence to be shown in section \ref{sec:turb}\ref{sec:condition}).

\subsection{Boundary layer development} \label{sec:evolve}

\begin{figure}
    \centerline{\includegraphics[width=0.9\linewidth]{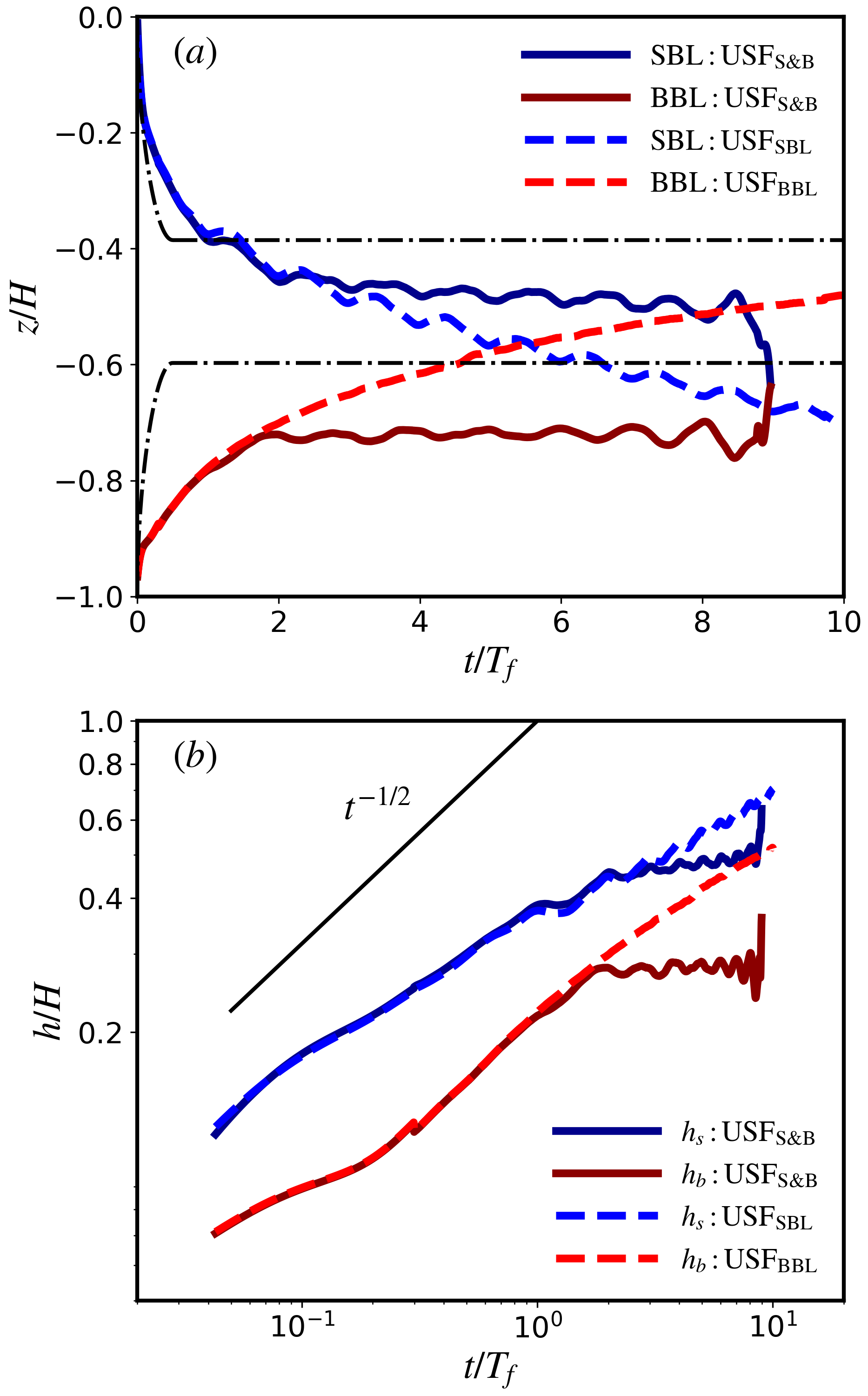}}
    \caption{Time history of (\textit{a}) the outer edge, and (\textit{b}) the thickness of the oceanic boundary layers for cases $\mathrm{USF_{S\&B}}$, $\mathrm{USF_{SBL}}$, and $\mathrm{USF_{BBL}}$. The SBL and BBL thicknesses are denoted as $h_s$ and $h_b$, respectively. The black dash-dotted lines show the depths of surface and bottom mixed layers predicted by \citet{Pollard1972GFD}.}
    \label{fig:mld}
\end{figure}

Figure \ref{fig:dtdz} shows the time history of the plane-averaged temperature gradient ${\mathrm{d}\langle \theta \rangle/\mathrm{d}z}$ for the three different cases considered here, i.e. $\mathrm{USF_{S\&B}}$, $\mathrm{USF_{SBL}}$, and $\mathrm{USF_{BBL}}$. For case $\mathrm{USF_{S\&B}}$ (figure \ref{fig:dtdz}\textit{a}), both the surface and bottom boundary layers develop under the combined effects of surface forcing and geostrophic current. The existence of the two separate thermoclines at early times is due to the entrainment sharpening of the adjacent density gradients for each of the SBL and BBL, as seen more persistently in the $\mathrm{USF_{SBL}}$ and $\mathrm{USF_{BBL}}$ cases. %Two strongly stable thermoclines, where the stratification becomes larger than the initial value, form immediately below the SBL and above the BBL at the initial stage. 
As the interior stratified layer gets thinner, these two thermoclines merge at $t/T_f \approx \frac{3}{2}$, and the temperature gradient of the resultant thermocline increases up to around 5 times of its initial value at $t/T_f \approx \frac{5}{4}$, compared to 3 times in $\mathrm{USF_{S\&B}}$ (figure \ref{fig:dtdz}\textit{b}) and 4 times in $\mathrm{USF_{BBL}}$ (figure \ref{fig:dtdz}\textit{c}). However, it is evidently observed that the strength of stratification in the thermocline for $\mathrm{USF_{S\&B}}$ and $\mathrm{USF_{SBL}}$ (figure \ref{fig:dtdz}\textit{a} and \textit{b}) oscillates with frequency $f$ before its ultimate disappearance, which is indicative of the modulation effect by inertial oscillations. The boundary layers in $\mathrm{USF_{S\&B}}$ cease to grow {once the two thermoclines merge}, and the only way for the system to evolve is to slowly mix the stratified fluids into the two boundary layers manifested by a gradual erosion of the interior stratification. This suggests that the merger between the SBL and BBL (figure \ref{fig:dtdz}\textit{a}) is not directly caused by boundary layer growth, but by a slow reduction in stratification within the thin layer between the two coexisting boundary layers. The interior stratification eventually disappears at $t/T_f \approx 9$ (i.e. the entire water column is in a neutrally stable condition), and the SBL and BBL merge into one fully developed boundary layer afterwards (consistent with that delineated in figure \ref{fig:ke}).

Figure \ref{fig:mld} shows the time history of the outer edges (\textit{a}) and thicknesses (\textit{b}) of the boundary layers for all 3 simulations. The boundary layer development for the isolated boundary layer scenarios (i.e. $\mathrm{USF_{SBL}}$ or $\mathrm{USF_{BBL}}$) is characterized by a rapid deepening, followed by a slow growth due to entrainment, similar to that observed by other authors for the upper \citep{Pham2017JGR} and bottom ocean Ekman layers \citep{Taylor2008b}. In contrast, the overlapping boundary layer ($\mathrm{USF_{S\&B}}$) flow evolves through three phases: a rapid deepening, an oscillating equilibrium, and a prompt merger. These three phases are separated by two important transitions: the merger of the two thermoclines separates the first two phses, and the disapearance of the internal stratification separates the final two phases (figure \ref{fig:dtdz}). The growths of SBL and BBL for $\mathrm{USF_{S\&B}}$ follow the isolated boundary layer cases (i.e. $\mathrm{USF_{SBL}}$ and $\mathrm{USF_{BBL}}$) in phase 1, but then depart in phase 2 starting at $t/T_f \approx \frac{3}{2}$, which coincides with the point where the two thermoclines merge into one stronger thermocline (see figure \ref{fig:dtdz}\textit{a}). This suggests that the merger of thermoclines marks an important moment at which the interaction between the two boundary layers becomes apparent. After the first transition, the SBL and BBL reach their quasi-equilibrium depths except with inertial oscillations superimposed on them. 
Once the stratification vanishes ($t/T_f \approx 9$), the depths of SBL and BBL change rapidly as there is no resistance to vertical mixing, resulting in the merger between the SBL and the BBL. Note that the main change after the first transition is more about boundary layer growth while the flow remains quasi-stationary, and the entire field only exhibit significant changes after the second transition (e.g. black solid line in figure \ref{fig:ke}).

\citet{Pollard1972GFD} predicts that the deepening of a constant-stress-driven Ekman layer into uniform stratification is given by,
\begin{equation} \label{eq:weight_func}
  h(t) = \left\{
    \begin{array}{ll}
      u_*\{4\left[ 1 - \cos{\left( ft \right)} \right]\}^{1/4}/\sqrt{N_0f}, 
      & 0 \le t/T_f \le 1/2 \\[2pt]
      2^{3/4}u_{*}/\sqrt{N_0f},         & t/T_f > 1/2.
    \end{array} \right.
\end{equation}
in which $h$ is the mixed layer depth, $u_{*}$ is the friction velocity, and $N_0$ is the background buoyancy frequency. In figure \ref{fig:mld}\textit{a}, The boundary layer developments based on the theory of \citet{Pollard1972GFD} are also included (black dash-dotted lines), using $u_{*s}$ and $u_{*b}$, respectively, as the velocity scale for the surface and bottom mixed layers. Here, $u_{*b}$ is estimated from the reduced form of the crosswise momentum equation \eqref{eq:momentum} when the flow reaches a quasi-equilibrium state, i.e. $u_{*b} = \left[ \overline{\langle u^{\prime}w^{\prime} \rangle}^2 + \overline{\langle v^{\prime}w^{\prime} \rangle}^2 \right]_{z={-H}}^{1/2}$ where $\overline{\langle u^{\prime}w^{\prime} \rangle}\vert_{z={-H}} = f \int_{-H}^0 \langle \overline{v} \rangle \mathrm{d}z$ and $\overline{\langle v^{\prime}w^{\prime} \rangle}\vert_{z={-H}} = f \int_{-H}^0 \left[ u_g - \langle \overline{u} \rangle \right] \mathrm{d}z$. Here, the values of $u_{*b}$ and $u_{*s}$ are very close ($u_{*b}/u_{*s}=1.04$) because we designed our simulation set-up to have comparable wind and current forcing conditions, which need not be the case in general. While \citet{Pollard1972GFD} excluded the late-time growth, the numerical studies of \citet{Jonker2013} and \citet{Pham2017JGR} predicted that the late-time growth is proportionate to $t^{1/2}$ regardless of the rotational effect. The BBL growth for case $\mathrm{USF_{S\&B}}$ (also $\mathrm{USF_{BBL}}$) indeed follows this $t^{1/2}$ relationship over the most part of phase 1 ($0.3<t/T_f<1.5$), but the SBL growth deviates from this relationship possibly due to the effect of Langmuir turbulence. In phase 2 ($1.5<t/T_f<9$), as the boundary layer turbulence continues to mix cooler water from below up into the SBL or warmer water from aloft down into the BBL, the background stratification continually change over time (figure \ref{fig:dtdz}). The increased stratification should lead to a slower boundary layer growth as is indeed observed in our simulations (figure \ref{fig:mld}). Additionally, the internal waves will perturb the boundary layers, and stress-driven mixed layers bounded by compliant (considered here) and rigid \citep{Pollard1972GFD} thermoclines are qualitatively different \citep{Chini2003JFM}. Hence, the boundary layer development shown here takes a different form from that predicted in the stress-driven Ekman layer \citep{Pollard1972GFD}.

In the following sections, we will focus mostly on the dynamics and structures in the overlapping boundary layers (i.e. case $\mathrm{USF_{S\&B}}$), given that they are much less explored compared to the scenario of the upper ocean SBL \citep[see][]{McWiliams:1997jfm, Sullivan2010ARFM, D'Asaro2014ARMS} or the current-driven BBL \citep[see][]{Taylor2007, Taylor2008b, trowbridge2018arms}. Results from $\mathrm{USF_{SBL}}$ and $\mathrm{USF_{BBL}}$ will still be used where necessary and serve as a reference to highlight the distinct features in the overlapping boundary layers. Since the flow in phase 1 behaves similar to the two isolated boundary layer counterparts, we will not discuss it further. Instead, we focus on LES solutions from the two separated inertial periods BM and AM to describe the flow features in phase 2 and phase 3. Also note that phase 2 should be very similar to phase 1, except that the boundary layer growth is stalled.

\section{Turbulence in the overlapping boundary layers} \label{sec:turb}
\subsection{Mean flow structure} \label{sec:mean}

\begin{figure*}
    \centerline{\includegraphics[width=1.0\linewidth]{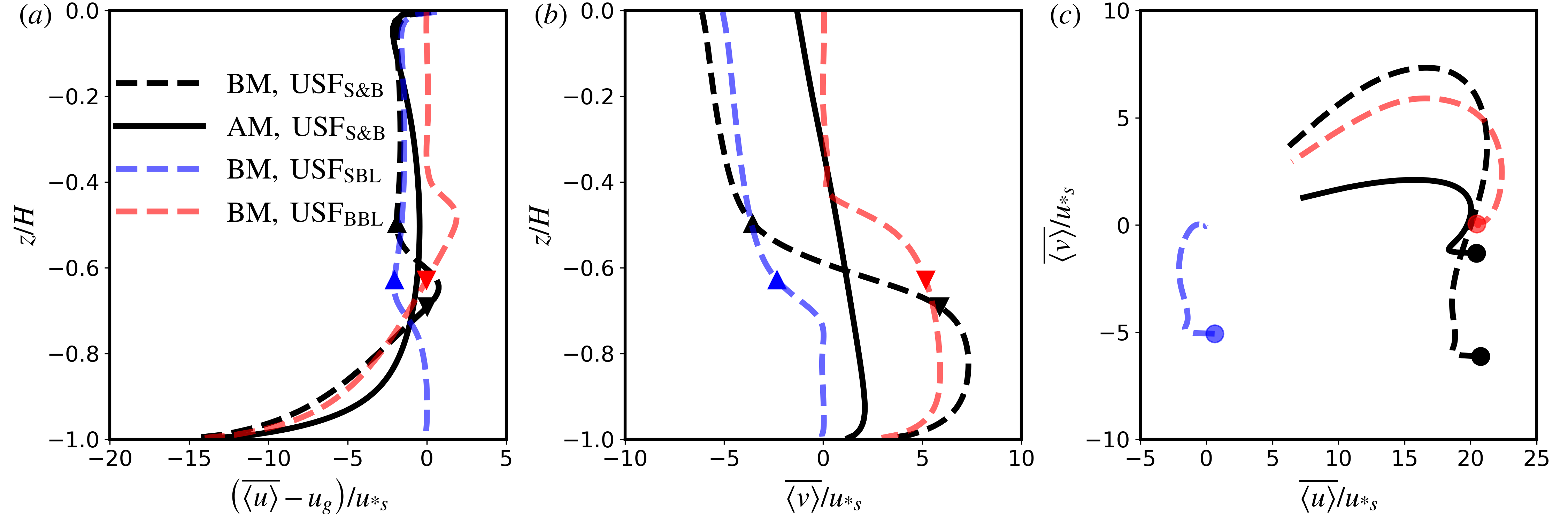}}
    \caption{Profiles of the mean velocity: (\textit{a}) streamwise ageostrophic component $(\overline{\langle u \rangle}-u_g)$ and (\textit{b}) crosswise component $\overline{\langle v \rangle}$ for case $\mathrm{USF_{S\&B}}$ before (black dashed line) and after (black solid line) the overlapping boundary layers merge, and (\textit{c}) Hodographs of the mean velocity vector, normalized by the surface friction velocity $u_{*s}$ (circles indicate values at the surface $z/H = 0$). The mean velocity components averaged over the inertial period BM for $\mathrm{USF_{SBL}}$ (blue dashed line) and $\mathrm{USF_{SBL}}$ (red) are also included.}
\label{fig:mean_stat}
\end{figure*}

Figure \ref{fig:mean_stat} shows the streamwise and crosswise components of the mean velocity for all three cases. Note that only the Eulerian velocity is shown here, while the Stokes drift is left out. On the grounds of dimensional analysis, four characteristic velocity scales matter in determining the flow regime, i.e. $u_{*s}$, $U_{s}$, $u_{*b}$, and $u_g$. The turbulence in the SBL scales with $u_{*s}$ and $U_{s}$, while the bottom turbulence scales with $u_{*b}$ and $u_g$. Because the forcing conditions are different among these cases, it is very difficult to find a universal velocity scale applicable for all three scenarios. Here, we are essentially comparing the absolute value of velocity-related statistics and use $u_{*s}=1.22\times10^{-2}\ \mathrm{m\ s^{-1}}$ as the scaling velocity.

Before the surface and bottom boundary layers merge (i.e. BM), as the interior temperature inversion still matters, the flow in the surface water behaves similar to that in Langmuir turbulence (case $\mathrm{USF_{SBL}}$, blue line), while the bottom water exhibits a similar flow pattern to the stratified bottom Ekman layer (case $\mathrm{USF_{BBL}}$, red line), similar to the flow field in phase 1 (figure \ref{fig:snapshot}\textit{c}). The overshoot in the thermocline for the downstream velocity (figure \ref{fig:mean_stat}) is inherent to the bottom Ekman layer flow \citep{Taylor2008b}. The magnitude of $\overline{\langle v \rangle}$ somewhat increases within the surface and bottom mixed layers relative to the isolated boundary layer counterparts, possibly because the two mixed layers are both confined to a shallower thickness due to stronger interior stratification (figure \ref{fig:dtdz}).

After the two boundary layers fully merge (i.e. AM), the profiles of $\overline{\langle u \rangle}$ and $\overline{\langle v \rangle}$ agree with the LES solutions of LSCs in \citet{Shrestha2019JFM} (case $\mathrm{C}_{2211}$ in figure 5 therein). The streamwise velocity is uniformly distributed in the central portion of the column, with most of the shear concentrated near the surface and bottom. It is interesting that $v$ is now mostly positive in the vertical column, suggesting that the BBL influence is stronger than the SBL influence. Unlike the wind-driven Langmuir turbulence in shallow water \citep{Tejada-Martinez:2007jfm, Kukulka2012jgr, Deng:2019JFM}, the overlapping boundary layer flow is also controlled by the bottom shear stress caused by the mean geostrophic current. As a result, the crosswise transport is mostly directed to the left of the $x-$direction in the vertical column except close to the surface (figure \ref{fig:mean_stat}\textit{b}). Because the crosswise velocity component in the SBL is pointing in the opposite direction to that in the BBL, they will counter-act each other when the SBL and BBL merge. Thus, the magnitude of $\overline{\langle v \rangle}$ is significantly reduced, and becomes nearly uniform in the vertical due to strong vertical mixing. The hodographs in figure \ref{fig:mean_stat}\textit{c} offer a different view of the mean horizontal velocity vector ($\overline{\langle u \rangle},\ \overline{\langle v \rangle}$). While cases $\mathrm{USF_{SBL}}$ and $\mathrm{USF_{BBL}}$ yield typical Ekman spirals in Langmuir turbulence and BBL turbulence respectively, the hodographs for $\mathrm{USF_{S\&B}}$ are very distorted due to the more complex behavior in the mean flow described above.

\subsection{Turbulence statistics} \label{sec:stat}

\begin{figure}
    \centerline{\includegraphics[width=0.8\linewidth]{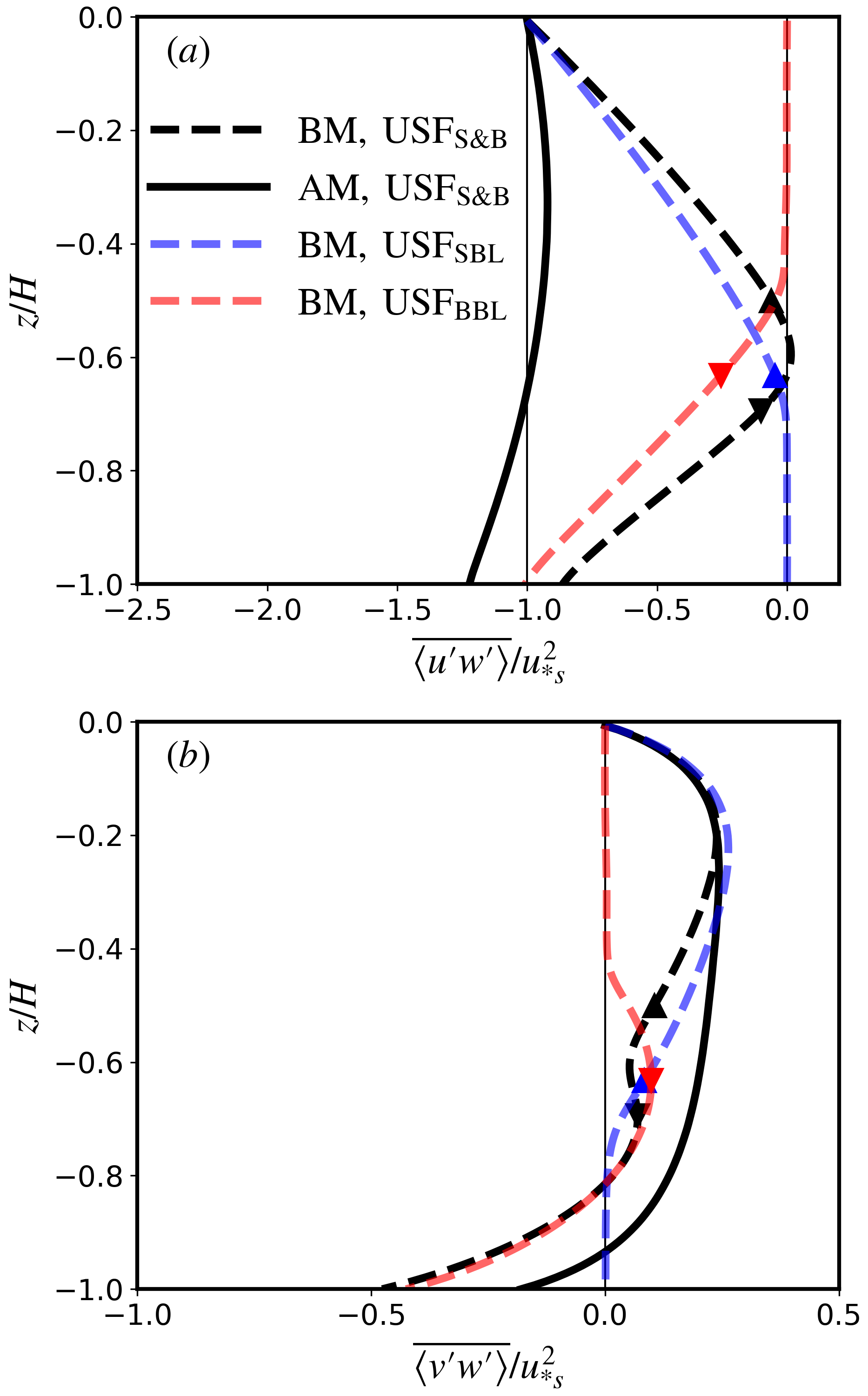}}
    \caption{Profiles of the mean vertical momentum flux: (\textit{a}) downstream component $\overline{\langle u^{\prime}w^{\prime} \rangle}$ and (\textit{b}) cross-stream  component $\overline{\langle v^{\prime}w^{\prime} \rangle}$ for case $\mathrm{USF_{S\&B}}$, normalized by the surface momentum flux $u_{*s}^2$. For legend, see the caption of figure \ref{fig:mean_stat}.}
\label{fig:mean_reynolds}
\end{figure}

Figure \ref{fig:mean_reynolds} shows the profiles of the vertical momentum flux, i.e. $\overline{\langle u^{\prime}w^{\prime} \rangle}$ and $\overline{\langle v^{\prime}w^{\prime} \rangle}$, in which both the resolved and SGS components are included. Continuity of the shear stress across the air-sea interface requires that $\overline{\langle u^{\prime}w^{\prime} \rangle}/u_{*s}^2 = 1$ at all times. Owing to the conservation of horizontal momentum, the distributions of $\overline{\langle u^{\prime}w^{\prime} \rangle}$ and $\overline{\langle v^{\prime}w^{\prime} \rangle}$ are closely related to the velocity profiles (figure \ref{fig:mean_stat}). When the flow reaches a quasi-steady state, the momemtum equation \eqref{eq:momentum} reduces to a balance among the turbulent stress divergence, pressure gradient force, and the Coriolis term, i.e.
\begin{subequations} \label{eq:reducedMom}
\begin{align} 
    {\partial \overline{\langle u^{\prime}w^{\prime} \rangle}}/{\partial z} & = f\langle \bar{v} \rangle
    \label{eq:streamwise}, \\[3pt]
    {\partial \overline{\langle v^{\prime}w^{\prime} \rangle}}/{\partial z} & = f\left(\overline{\langle{u}\rangle}+u_{s}-u_g \right)
\end{align}
\end{subequations}
Before the merger, the flux of streamwise momentum approaches its minimum magnitude $\overline{\langle u^{\prime}w^{\prime} \rangle} \approx 0$ at the depth $z/H \approx 0.6$ (figure \ref{fig:mean_reynolds}\textit{a}), where the local flux gradient ${\partial \overline{\langle u^{\prime}w^{\prime} \rangle}}/{\partial z} |_{z/H \approx 0.6} = 0$ and thus the crosswise velocity $\overline{\langle {v} \rangle}$ changes its sign (black dashed line in figure \ref{fig:mean_stat}). The magnitudes of $\overline{\langle u^{\prime}w^{\prime} \rangle}$ and $\overline{\langle v^{\prime}w^{\prime} \rangle}$ are reduced within the boundary layers compared to that in the isolated boundary layer scenarios (blue and red dashed lines) due to the differences in boundary layer depths. Note that $\overline{\langle v^{\prime}w^{\prime} \rangle}$ exhibits a nonzero value in the stratified layer, suggesting that the SBL and BBL are in a partly communicating regime. While the interior stratification still inhibits the vertical mixing of the entire water column, this dynamical coupling is potentially enabled by the internal waves generated thereabout due to the interaction of boundary layer turbulence with the interior stratification, which will be described in section \ref{sec:IGW}. After the merger, the bottom stress increases by about 50\% and now it is greater than the surface value ($u_{*b}/u_{*s}=0.99$ before the merger and 1.24 after merger). This leads to a positive crosswise velocity $\langle \bar{v} \rangle > 0$ almost over the entire water column except near the surface as shown in figure \ref{fig:mean_stat}\textit{b}. The enhanced bottom stress (after the merger) is caused by the penetration of Langmuir circulations down to the bottom wall (i.e. forming the so-called LSCs, see section \ref{sec:condition}), which has potential implications for coastal sedimentation and erosion \citep{Gargett2004Science}.

\begin{figure*}
    \centerline{\includegraphics[width=1.0\linewidth]{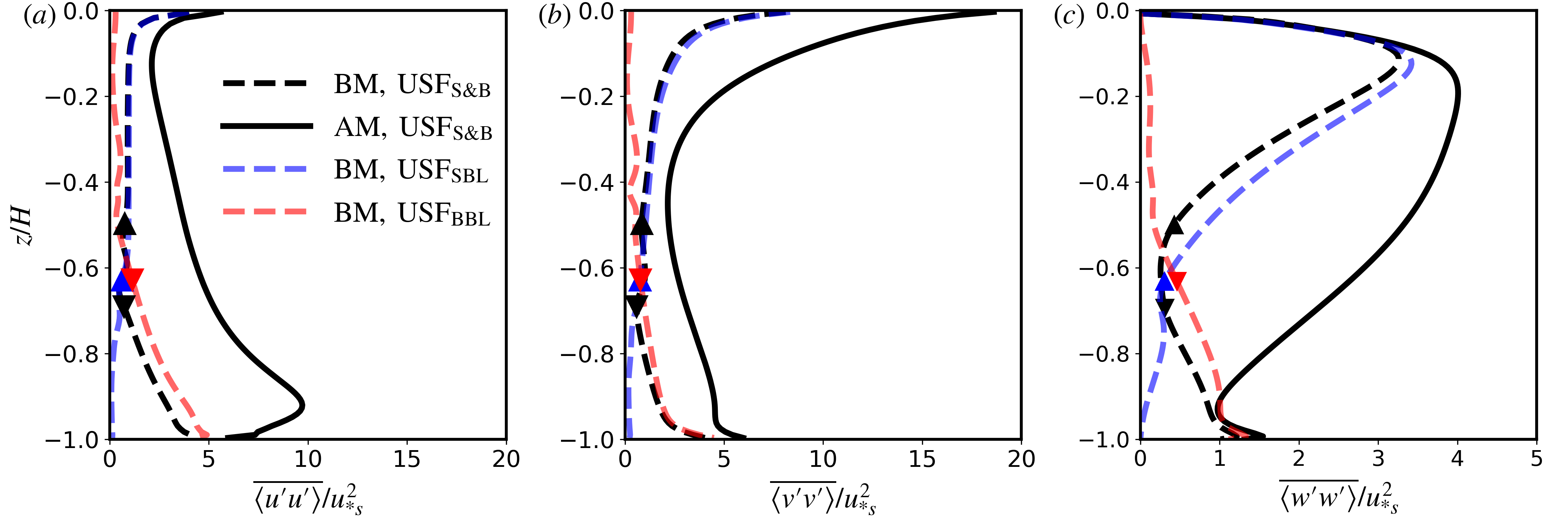}}
    \caption{Profiles of the mean velocity variances for the resolved motion: (\textit{a}) $\overline{\langle u^{\prime}u^{\prime} \rangle}$, (\textit{b}) $\overline{\langle v^{\prime}v^{\prime} \rangle}$, and (\textit{c}) $\overline{\langle w^{\prime}w^{\prime} \rangle}$ for case $\mathrm{USF_{S\&B}}$, normalized by the surface momentum flux $u_{*s}^2$. For legend, see the caption of figure \ref{fig:mean_stat}.}
\label{fig:mean_variance}
\end{figure*}

The turbulent intensities also exhibit remarkable changes in magnitude when the flow turns into a fully merged boundary layer (see figure \ref{fig:mean_variance}), partly because the stronger bottom shear (figure \ref{fig:mean_stat}) and the enhanced bottom stress (figure \ref{fig:mean_reynolds}) promote turbulent mixing in the vertical column. Before the merger, all the three components appear to be an amalgamation of turbulence intensities in $\mathrm{USF_{BBL}}$ and $\mathrm{USF_{BBL}}$ (blue and red dashed lines), suggesting that the interaction between the SBL and BBL is not very strong. The vertical turbulent intensities $\overline{\langle w^{\prime}w^{\prime} \rangle}$ are nonzero for all three cases (dashed lines in figure \ref{fig:mean_variance}\textit{c}) in the stably stratified layer, which could be kinetic energy carried by internal waves radiating away into the stratified layer. After the two boundary layers are merged, based on the shape of the vertical profiles, we infer that the streamwise component $\overline{\langle u^{\prime}u^{\prime} \rangle}$ is dominated by shear production at the bottom, while the crosswise $\overline{\langle v^{\prime}v^{\prime} \rangle}$ and vertical $\overline{\langle w^{\prime}w^{\prime} \rangle}$ components are mainly dominated by the surface forcing associated with Langmuir turbulence in the upper layer. It should be noted that $\overline{\langle u^{\prime}u^{\prime} \rangle}$ and $\overline{\langle v^{\prime}v^{\prime} \rangle}$ from case $\mathrm{USF_{BBL}}$ (red line) both approach a small nonzero value above the bottom mixed layer because of the sampling error involved in filtering out the inertial oscillations associated with the horizontal current. As the flow in $\mathrm{USF_{BBL}}$ is almost non-turbulent above the BBL, it takes more time (i.e. more than an inertial period) for the current-driven flow to bounce back to an equilibrium solution due to strong inertia and no assistance from turbulent mixing. Nevertheless, the sampling error for case $\mathrm{USF_{S\&B}}$ should be very small because the highly turbulent entrainment of the narrow stratified region will greatly reduce the effect of inertial acceleration.

\subsection{Comparison with Langmuir supercells}\label{sec:condition}

The upper ocean flow is populated with the well-known Langmuir circulations, which are generated by wave-current interactions via the CL2 instability, i.e. the wave-induced Stokes drift shear tilts the vertical vorticity (associated with the crosswind shear) into the downwind direction, forming pairs of couter-rotating vortices \citep{Leibovich1983arfm}. In shallow-water regions ($\sim$ 15 m), as the Stokes drift velocity persists even at the seabed, Langmuir circulations occupy the entire water column (i.e. the so-called Langmuir supercells), with a lateral scale 3$\sim$6 times the water depth \citep{Gargett2004Science, Tejada-Martinez:2007jfm}. The interactions of the bottom shear with the surface waves will also affect the morphology and characteristics of the Langmuir supercells \citep{Kukulka2011jgr}.

In an intermediate-depth ocean where the Stokes drift velocity vanishes below certain vertical level, one interesting question is how Langmuir circulations behave when the two boundary layers are fully merged. Will these coherent circulations still be confined to the upper half of the water column, or will they also extend towards the sea bottom? Visual evidence in figure \ref{fig:snapshot_xy} seems to suggest the latter. Here, we use a conditional sampling method for the LES solutions to educe the size and strength of Langmuir structures. Based on the preconception that Langmuir circulations induce strong downwelling motions, the conditional sampling operation for any physical quantity $\phi$ is defined as,
\begin{equation} \label{condition}
\begin{split}
    \mathring{\phi}(x_r, y_r, & x',y',z',t) =\left \langle \phi(x_r+x',y_r+y',z',t) \big| \mathscr{E}\right \rangle, \\
    & \mathrm{as}\ \mathscr{E}:\ w(x_r,y_r,z_{*},t) \leq - \overline{\langle w^{\prime}w^{\prime} \rangle}^{1/2} \big|_{max},
\end{split}
\end{equation}
in which $(x_r, y_r)$ is the reference point (that enumerates all the grid point on the $x-y$ plane) with $(x',y')$ being the distance from $(x_r, y_r)$ in the horizontal direction, and $z_*$ is the depth at which $\overline{\langle w^{\prime}w^{\prime} \rangle}^{1/2}$ attains its maximum value $\overline{\langle w^{\prime}w^{\prime} \rangle}^{1/2} \big|_{max}$ \citep{McWiliams:1997jfm}. Alternative definitions of the conditional event $\mathscr{E}$ have been used. such as one based on upwelling motions, but they do not yield a flow structure very different from the one reported here in terms of the size and strength. 

It is worth noting that in previous LES studies of shallow-water Langmuir turbulence (where wind and waves are co-aligned in the streamwise direction) \citep{Tejada-Martinez:2007jfm, Kukulka2012jgr, Deng:2019JFM}, the Langmuir supercell structures are normally distilled from the LES field as the streamwise-averaged turbulent fluctuations. In those studies, Langmuir circulations are roughly aligned with the wind and wave directions as the Coriolis rotation is usually omitted \citep{Grosch2016jpo}. However, with the inclusion of the Earth's rotation (as we have considered in this work), the orientation of Langmuir circulations is somewhat deflected from the wind direction and also changes with increasing depth \citep{McWiliams:1997jfm}.

\begin{figure*}
    \centerline{\includegraphics[width=0.9\linewidth]{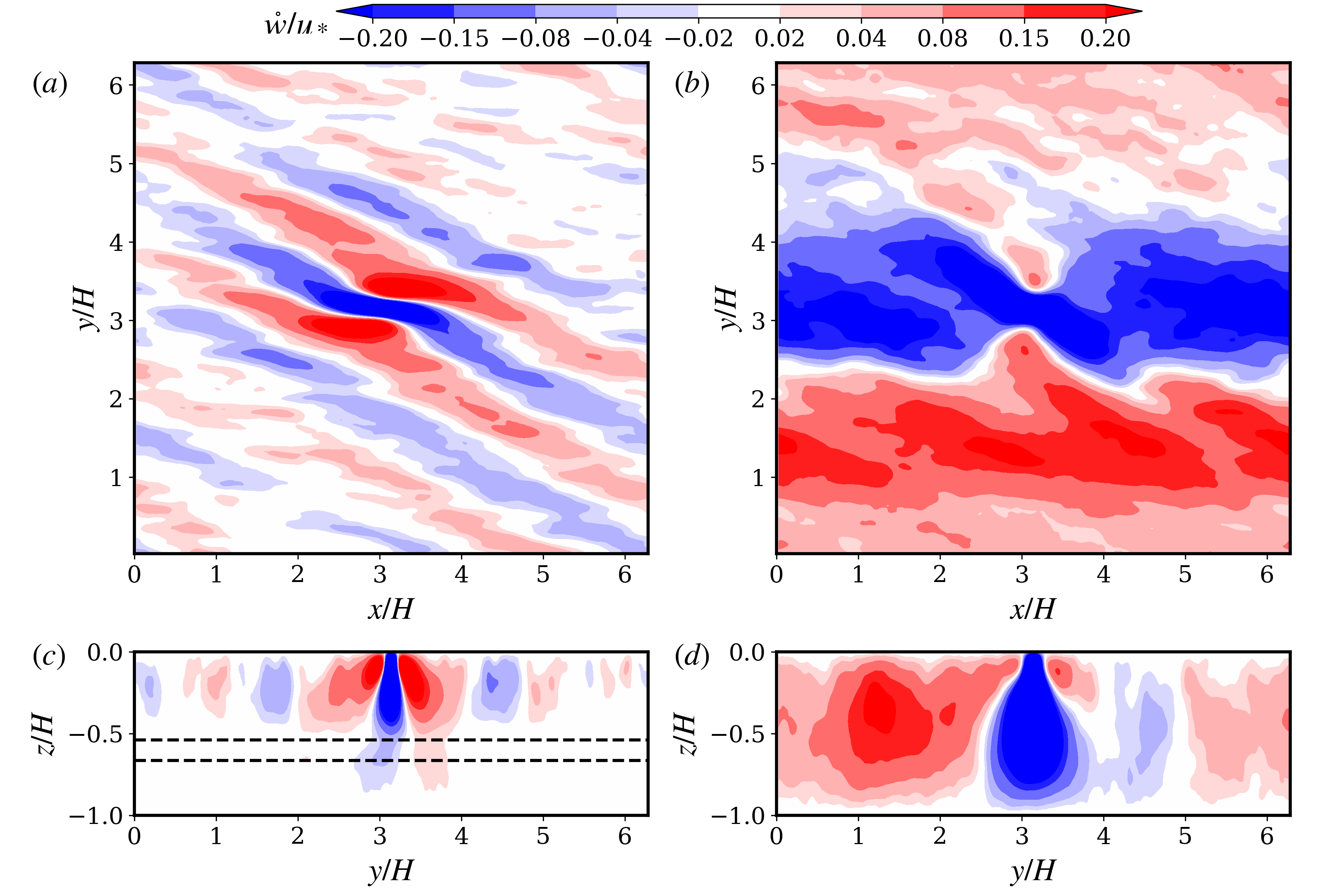}}
    \caption{Conditional-averaged vertical velocity $\mathring{w}/u_{*}$ in the $x-y$ plane ($z/H=-0.2$) and $y-z$ plane ($x/H=\pi$) before and after the merger for the simulation $\mathrm{USF_{S\&B}}$. Panels $\textit{a}$ and $\textit{c}$ show $\mathring{w}/u_{*}$ before the merger; Panels $\textit{b}$ and $\textit{d}$ show $\mathring{w}/u_{*}$ after the merger. The two dashed lines indicate the outer edges of the upper and lower mixed layers based on the definition \eqref{eq:mld}.}
\label{fig:wcond}
\end{figure*}

Figure \ref{fig:wcond} shows the contour plots of $\mathring{w}/u_{*}$ in the $x-y$ plane ($z/H=-0.2$) and $y-z$ plane ($x/H=\pi$) from simulation $\mathrm{USF_{S\&B}}$ as noted in the caption. Before the merger, Langmuir circulations are mainly confined to the SBL, but occasionally induce upwelling and downwelling motions in the stratified layer (figure \ref{fig:wcond}\textit{c}), which is likely the main source of internal waves there \citep{Chini2003JFM}. The Langmuir cells are elongated in the longitudinal direction, with the axis oriented slightly to the right of the wind direction (i.e. positive $x-$direction) because of the Ekman shear (figure \ref{fig:mean_stat}). The cell pattern appears antisymmetric about the conditioning origin, i.e. $(x_r=\pi,y_r=\pi)$, with a lateral span of 2 times their vertical extension ($\sim 0.5H$), consistent with that described in \cite{McWiliams:1997jfm}.

However, the size and orientation of Langmuir circulations become distinctively different after the merger. Interestingly, the Langmuir circulations at this flow stage extend down to the sea bottom, even though the Stokes drift velocity is zero in the lower half of the water column (see figure \ref{fig:domain}\textit{b}). 
%This is possibly because the current shear near the bottom (figure \ref{fig:mean_stat}) can also tilts the vertical vorticity associated with the random crosswise shear into the streamwise direction in a way similar to the CL2 instability, forming pairs of couter-rotating vortices in the lower layer.
Based on the time history of $\langle w^{\prime}w^{\prime} \rangle$ (not shown), the full-depth Langmuir circulations emerge over a very short period of time $9<t/T_f<9.2$. They are oriented to the right of the wind direction at first, and then they are adjusted to be aligned with the wind direction (lower panels in figure \ref{fig:snapshot_xy}), see the supplementary movie. When the flow finally reaches an equilibrium state, the whole simulation domain now resolves only one pair of counter-rotating Langmuir rolls, with a lateral scale of about 6 times the water depth (figure \ref{fig:wcond}\textit{b} and \textit{d}). %Although to what extent that the horizontal periodicity and the limited domain size matter is up to debate, it is clear that the the lateral extension of the Langmuir circulations is significantly enlarged after the two boundary layers fully merge.
Clearly the simulation domain is too small and the final lateral and streamwise patterns are impacted by the use of periodic boundary conditions. Nevertheless, we can safely conclude that after the two boundary layer merge, the Langmuir circulations occupy the entire water column and their lateral extension is much larger than before the merger. The full-depth Langmuir cells become less distorted and appear more aligned in the streamwise direction, possibly due to weaker crosswise current shear (figure \ref{fig:mean_stat}\textit{b}) as reported in \cite{Kukulka2011jgr}. 

Relative to the central downwelling region, the full-depth Langmuir cells exhibit stronger upwelling motions on the right flank compared to the left flank (facing downstream), which could change depending on the forcing conditions. The potential impact of varying wind-wave-current conditions on the resulting appearance of Langmuir structures is out of the scope here, but should be explored in the future. 
%We expect that the strength and orientation of the full-depth Langmuir cells should be dependent on the magnitude of the geostrophic current, as the Eulerian mean shear and bottom-generated turbulence are all influential factors affecting the generation of Langmuir circulations. Observations suggest that the strength of LSCs is negatively modulated by increasing current \citep{Kukulka2011jgr}. The potential impact of varying wind-wave-current conditions on the resulting appearance of Langmuir structures is out of the scope here, but should be explored in the future. Still, the pattern transition of Langmuir circulations presented here could serve as a guidance for parameterizing the vertical mixing due to Langmuir turbulence in coastal regions.
According to \cite{Shrestha2019EFM}, the upwelling and downwelling motions associated with the full-depth Langmuir circulations will induce a phase-locked modulation on the bottom stress, which leads to elevated bottom stresses seen in figure \ref{fig:mean_reynolds} (black solid lines). This is the main cause for the increased streamwise turbulent stress $\overline{\langle u^{\prime}w^{\prime} \rangle}_b$ and also total stress $u_{*b}^2$ at the bottom (in terms of magnitude) after the merger, see figure \ref{fig:mean_reynolds}. The reduction in the magnitude of crosswise turbulent stress $\overline{\langle v^{\prime}w^{\prime} \rangle}_b$ near the bottom is attributed to the counterbalance of momentum transfer driven by the surface-forcing and bottom-shear mechanisms.

% --
\subsection{Turbulent kinetic energy budget} \label{sec:budget}

To better understand the energy transport in the vertical column, we examine the contributions from various production and destruction terms in the turbulent kinetic energy (TKE) budget. The resolved kinetic energy averaged over an inertial period $K=\frac{1}{2}\overline{\left<{u_i u_i}\right>}$ can be split into, 
\begin{equation} \label{eq:totalKE}
    K=\frac{1}{2}\overline{\left<{u_i u_i}\right>} = 
    \underbrace{ \vphantom{ \left<\overline{u_i'u_i'}\right>_y }
    \frac{1}{2}\overline{\left<{u_i}\right>}\ \overline{\left<{u_i}\right>} 
    }_{\mathrm{MKE}} +
    \underbrace{ \vphantom{ \left<\overline{u_i'u_i'}\right>_y }
    \frac{1}{2}\overline{\left<{u_i}\right>^{\prime\prime}\left<{u_i}\right>^{\prime\prime}}
    }_{\mathrm{IOKE}} +
    \underbrace{\frac{1}{2}\overline{\left<u_i'u_i'\right>}}_{\mathrm{TKE}}
\end{equation}
Here, the double prime denotes the temporal fluctuation. The first term on the RHS of \eqref{eq:totalKE} is the mean kinetic energy (MKE), the second term represents the kinetic energy in the inertial oscillations (IOKE), and the third term is the time-averaged TKE. Under horizontally homogeneous conditions, the temporal evolution of the resolved-scale TKE ($k=\langle u^{\prime}_i u^{\prime}_i\rangle/2$) is given by,
\begin{equation}  \label{eq:TKE}
\begin{split}
    \frac{\partial{k}}{\partial{t}} &= {\underbrace{-\left[\langle u_i^{\prime}w^{\prime}\rangle + \langle \tau^{d}_{i3} \rangle \right] \frac{\mathrm{d}\langle u_i\rangle }{\mathrm{d}z}}_{P_k}} 
    {\underbrace{-\langle u^{\prime}w^{\prime}\rangle \frac{\mathrm{d}{u_{s}}}{\mathrm{d}z}}_{S_k}}
    +\underbrace{\vphantom{ \frac{\mathrm{d}{u_{s}}}{\mathrm{d}z} }
    \alpha g\langle w^{\prime}\theta^{\prime}\rangle}_{B_k} \\
    & \quad 
    \underbrace{\vphantom{\frac{\partial u^{\prime}_i}{\partial{x_j}}}
    -\frac{1}{2}\frac{\mathrm{d}\langle u^{\prime}_iu^{\prime}_iw^{\prime}\rangle }{\mathrm{d}z} }_{T_k}
    \underbrace{\vphantom{\frac{\partial u^{\prime}_i}{\partial{x_j}}}
    -\frac{1}{\rho_0}\frac{\mathrm{d}\langle w^{\prime}p^{\prime}\rangle }{\mathrm{d}z}}_{\Pi_k}
    +\underbrace{\vphantom{\frac{\partial u^{\prime}_i}{\partial{x_j}}} 
    \frac{\mathrm{d}\langle u^{\prime}_i\tau^{d\prime}_{i3}\rangle}{\mathrm{d}z}}_{D_k}
    \underbrace{-\langle \tau^{d}_{ij}\frac{\partial u_i}{\partial{x_j}}\rangle}_{\epsilon}
\end{split}
\end{equation}
The terms on the RHS of \eqref{eq:TKE} are identified as shear production $P_k$, Stokes production $S_k$, buoyancy production $B_k$, turbulent transport $T_k$, pressure transport $\Pi_k$, SGS diffusion $D_k$, and SGS dissipation rate $\epsilon$ (assumed to be a good proxy for TKE dissipation rate), respectively. Because the filter scale is much larger than the Kolmogorov scale, the viscous dissipation of resolved TKE is negligible.

\begin{figure*}
    \centerline{\includegraphics[width=0.9\linewidth]{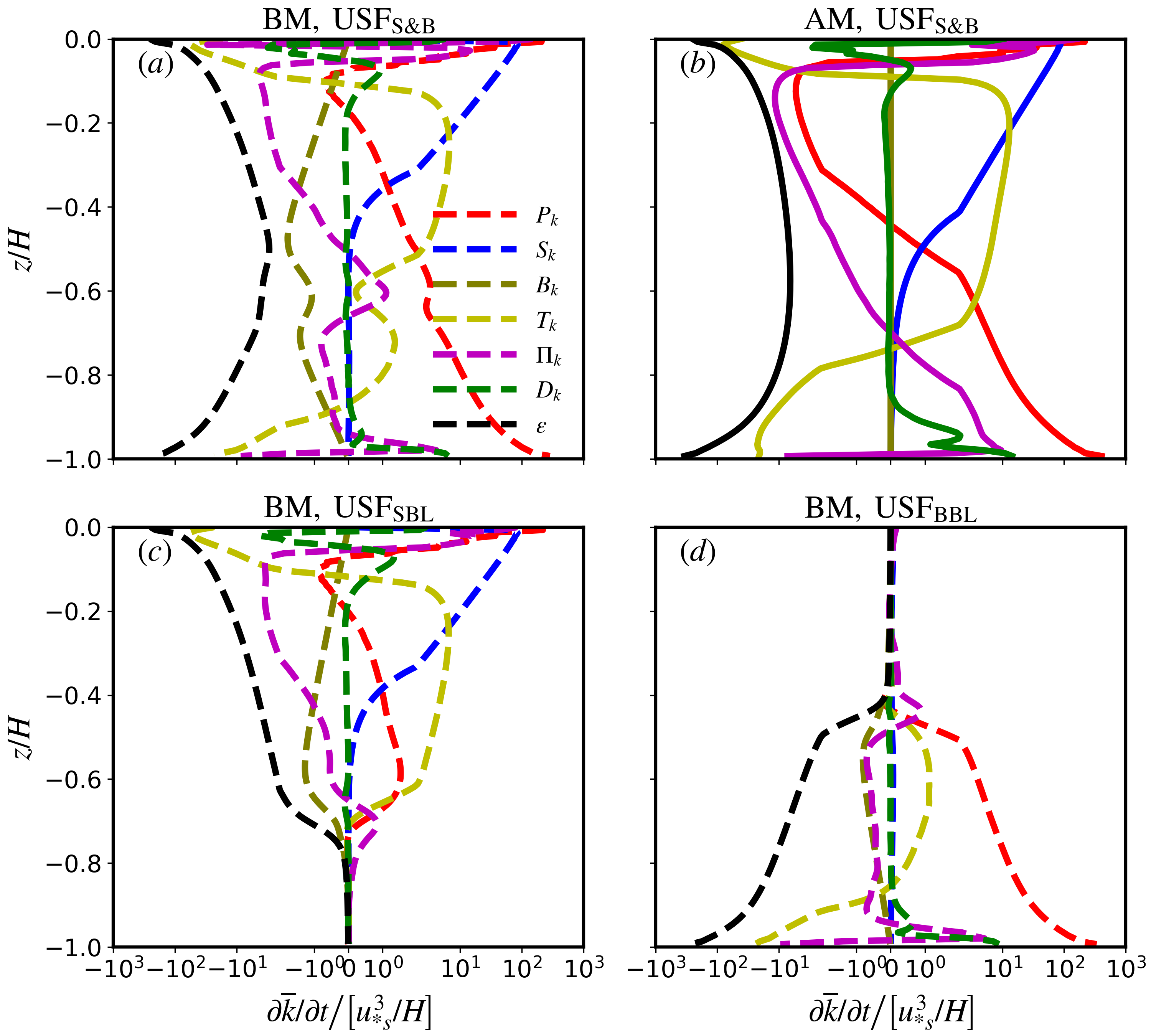}}
    \caption{Vertical variations of terms in the TKE budget equation \eqref{eq:TKE} for all 3 cases (\textit{a}) $\mathrm{USF_{S\&B}}$ BM, (\textit{b}) $\mathrm{USF_{S\&B}}$ AM, (c) $\mathrm{USF_{SBL}}$, and (c) $\mathrm{USF_{BBL}}$. Each term is scaled by $u_{*s}^3/H$.}\label{fig:budget}
\end{figure*}

Figure \ref{fig:budget} shows the time-averaged terms in \eqref{eq:TKE} before and after the merger for case $\mathrm{USF_{S\&B}}$, normalized by $u_{*s}^3/H$. For comparison, results from simulations $\mathrm{USF_{SBL}}$ and $\mathrm{USF_{BBL}}$ are also included. The TKE budget terms for case $\mathrm{USF_{BBL}}$ (figure \ref{fig:budget}\textit{d}) are in good agreement with that of the bottom Ekman layer in \cite{Taylor2008b} (figure 11 therein), thus lending more confidence to the fidelity of the present model. As \cite{Taylor2007} pointed out, the pressure transport $\Pi_k$ becomes the major source term in the pycnocline ($z/H \approx -0.4$ in figure \ref{fig:budget}d), which implicates the generation of internal waves by BBL turbulence. This is also true for the isolated SBL scenario as $\Pi_k$ is positive in the pycnocline at $z/H \approx -0.7$ (figure \ref{fig:budget}\textit{c}), suggesting that internal waves are also generated by the interaction of Langmuir turbulence and stratification. The energy budget in figure \ref{fig:budget}\textit{c} is also consistent with typical Langmuir turbulence in deep ocean \citep{Grant2009JPO}. The shear production is very small in the upper portion of the boundary layer as the Stokes production plays a dominant role in the generation of Langmuir turbulence.

For case $\mathrm{USF_{S\&B}}$, the energy budget terms before the merger appear to be an amalgamation of those for the isolated boundary layer cases. The production and dissipation terms are mostly concentrated near the surface and bottom where the mean current shear is strong. Before the merger, the production $P_k$ and $S_k$ are the primary source of TKE to balance dissipation $\epsilon$ near the surface and bottom, while $B_k$ only accounts for a negligibly small fraction for TKE budget. The shear production is non-zero in the stratified layer due to small but nonzero $\overline{\langle v^{\prime}w^{\prime} \rangle}$ (figure \ref{fig:mean_reynolds}\textit{b}) and local enhanced shear (figure \ref{fig:mean_stat}). The turbulent transport $T_k$ acts as a sink near the surface and bottom, and serve as a source in the bulk of the two boundary layers. Because $T_k$ represents the non-local transport contribution to TKE, this suggests that kinetic energy is transferred from the surface and bottom layers towards the interior of the boundary layers via non-local transport mechanisms. $T_k$ is approxiamtely zero at the interface between the SBL and BBL, implying that the non-local transport is primarily confined to the boundary layers. The pressure transport $\Pi_k$ is positive in the pycnocline at $z/H \approx -0.6$, and it serves as a primary sink term in the SBL ($-0.5 < z/H < -0.05$) and acts as a secondary sink in the BBL ($-0.95 < z/H < -0.7$) in terms of the magnitude. This is another evidence that suggests the presence of internal waves, and links their energy source to the boundary layer turbulence with larger contributions originating from Langmuir turbulence in the SBL. As the internal waves are generated in the pycnocline, the change of sign for $\Pi_k$ in the vertical (i.e. $z/H \approx -0.5$ and $-0.65$) indicates that the vertical energy flux $\langle p^{\prime}w^{\prime}\rangle$ is radiated away (upward and downward) from the BBL and SBL. The ocean surface and bottom pose a natural barrier on the vertical propagation of internal waves, thus $\Pi_k$ changes sign at $z/H \approx -0.05$ and $-0.95$ and acts as a sink term near the surface and bottom regions (i.e. $z/H < -0.95$ and $z/H > -0.05$). However, $\Pi_k$ is much smaller than the dissipation $\epsilon$, suggesting that the energy loss associated with internal waves is very small compared to the total dissipated energy. Even though the energy carried away by internal waves is small, the waves clearly impact the boundary layer structure (figure \ref{fig:snapshot_xy}\textit{d}) and may exert a significant influence on the evolution of background potential energy \citep{Taylor2007}.

After the overlapping boundary layers fully merge (figure \ref{fig:budget}\textit{b}), $P_k$ and $\epsilon$ are further enhanced near the seabed, owing to greater current shear near the bottom (see figure \ref{fig:mean_stat}\textit{a}). In the upper portion of the boundary layer, turbulence is energized by Stokes production $S_k$ and even loses energy to MKE as $P_k<0$.%; while the lower boundary layer turbulence production is mainly attributed to the current shear near the bottom ($u_s$ is nearly zero for this depth range). 
Because the magnitude of $\overline{\langle u^{\prime}w^{\prime} \rangle}$ increases (figure \ref{fig:mean_reynolds}\textit{a}), the Stokes production $S_k$ also becomes larger. As $\epsilon$ does not change much in the surface layer, the increase of $S_k$ leads to negative $P_k$ at some levels $0.05 < z/H < 0.3$. $\Pi_k$ and $T_k$ now have opposite signs (e.g. $T_k$ is negative near the surface and bottom, and positive in the central column), suggesting that boundary layer turbulence transports energy from the surface and bottom turbulent motions to the fluid in the central part ($-0.7 < z/H < -0.1$), while the pressure transport redistribute energy in the vertical by transferring energy in the central region to the surface and bottom layers.

\begin{figure*}
    \centerline{\includegraphics[width=0.8\linewidth]{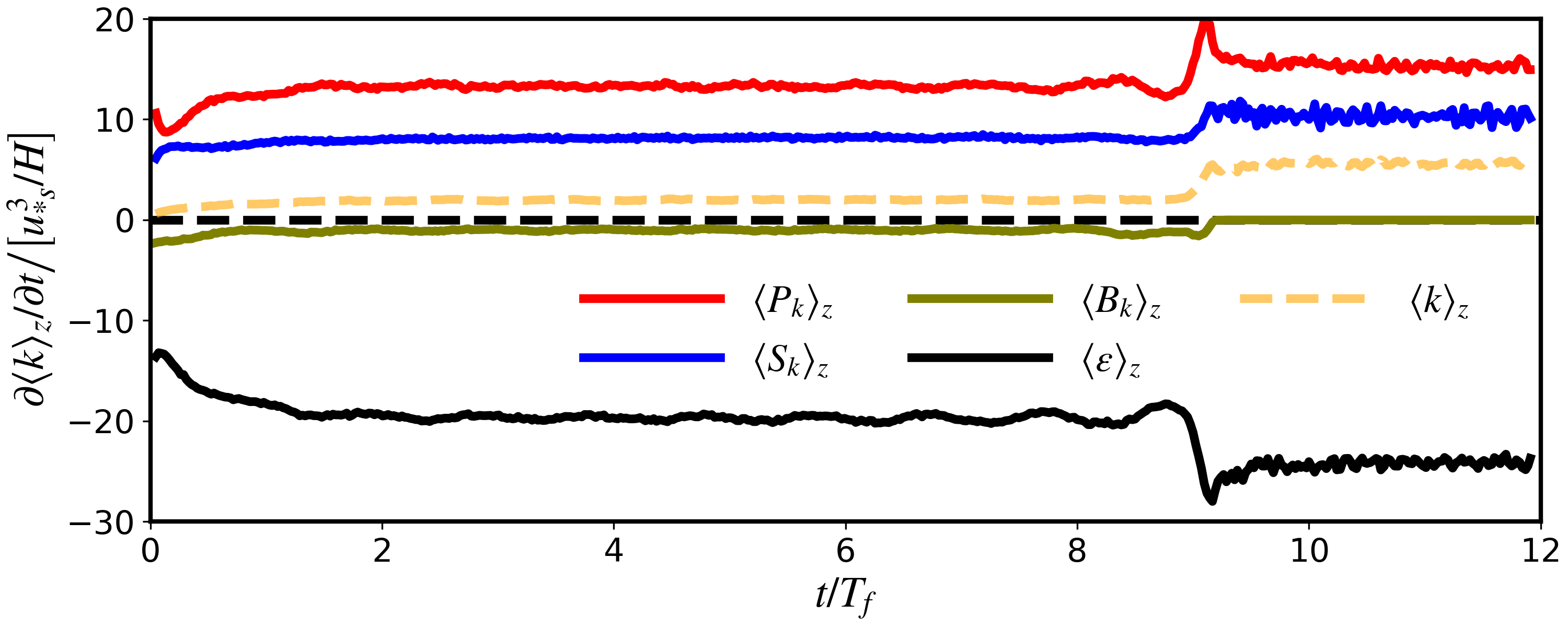}}
    \caption{Time history of the depth-averaged terms in TKE budget equation \eqref{eq:TKE} for case $\mathrm{USF_{S\&B}}$, normalized by $u_{*s}^3/H$. The transport terms in equation \eqref{eq:TKE} would integrate to zero and thus they are not included here.}
    \label{fig:budgetT}
\end{figure*}

To explain the augmentation of turbulence levels after the overlapping boundary layers fully merge (figure \ref{fig:mean_variance}), the production and destruction terms in \eqref{eq:TKE} are further integrated in the vertical direction (from the surface to the bottom). Figure \ref{fig:budgetT} shows the time history of the depth-averaged terms in \eqref{eq:TKE} for $\mathrm{USF_{S\&B}}$, denoted by $\langle \cdot \rangle_z$ (e.g. $\langle {P_k} \rangle_z = \int_{-H}^{0} P_k \mathrm{d}z \left/ H \right.$). Note that the depth-averaged transport terms, e.g. $\langle {T_k} \rangle_z$, $\langle {D_k} \rangle_z$, and $\langle {\Pi_k} \rangle_z$, are not shown because they should be identically zero. The production terms $\langle {P_k} \rangle_z$ and $\langle {S_k} \rangle_z$ are balanced to within a few percent by the dissipation $\langle {\epsilon} \rangle_z$. %The magnitude of the buoyancy term $\langle B_k \rangle_z$ is much smaller than $\langle {\epsilon} \rangle_z$, but the interior stratification plays an impressive role in hampering the vertical mixing. 
The enhancement of TKE $\langle k \rangle_z$ after the merger is attributed to the increased $\langle {P_k} \rangle_z$ and $\langle {S_k} \rangle_z$. Since ${P_k}$ acts as a sink term in the upper ocean, the increase of $\langle {P_k} \rangle_z$ is mainly caused by the shear production associated with the increased bottom shear, which arises from the full-depth Langmuir circulations that modulate the BBL dynamics. The full-depth Langmuir circulations also promote the vertical momentum transfer, leading to larger Stokes production that causes transfer of wave energy to a deeper depth.

% --
\section{Internal waves and turbulence} \label{sec:IGW}

The visual evidence presented above clearly confirms the presence of internal waves within the stably stratified layer. However, tracing the origin and evaluating their dissipation remain elusive due to cascades of nonlinear interactions \citep{Garrett1979ARFM, Staquet:2002ARFM}. The internal wave dynamics are strongly dependent on the vertical density structure of the water column \citep{Massel:2015}. For the overlapping boundary layers $\mathrm{USF_{S\&B}}$, the vertical density distribution exhibits a three-layer structure where the surface and bottom uniform layers are separated by a non-uniform layer in between. As the buoyancy frequency is time-varying, it is difficult to obtain a closed analytical solution to this problem.

The two dominant restoring forces, which determine the existence of internal waves, are the vertical stratification (with buoyancy frequency $N$) and Earth's rotation (with inertial frequency $f$). These two factors force fluid parcels to oscillate back and forth about their equilibrium positions. For clarity, internal waves dominated by the buoyancy force are called internal gravity waves, while those mainly affected by Coriolis force are called inertial waves. The associated internal waves are characterized by the dispersion relation below \citep{Phillips:1977},
\begin{equation} \label{eq:IGW_freq}
    \sigma^2 = (N^2k^2 + f^2m^2)/(k^2 + m^2) = N^2\cos^2{\gamma} + f^2\sin^2{\gamma}
\end{equation}
in which $\sigma$ denotes the internal wave frequency, $k$ and $m$ are the vertical and horizontal wave numbers respectively, and $\gamma$ is the angle between the wave vector and horizontal plane. Therefore, internal waves span the frequency range between the inertial frequency $f$ and the buoyancy frequency $N$. Since the internal wave periods are much longer than those of surface waves, the Stokes drift induced by internal waves is negligibly small. Therefore, the internal waves are well resolved here, rather than being filtered like the surface waves in deriving wave-averaged equations \eqref{eq:continuity} to \eqref{eq:temperature}.

% --
\subsection{Modulation of heat transfer} \label{sec:heatFlux}

\begin{figure*}
    \centerline{\includegraphics[width=0.8\linewidth]{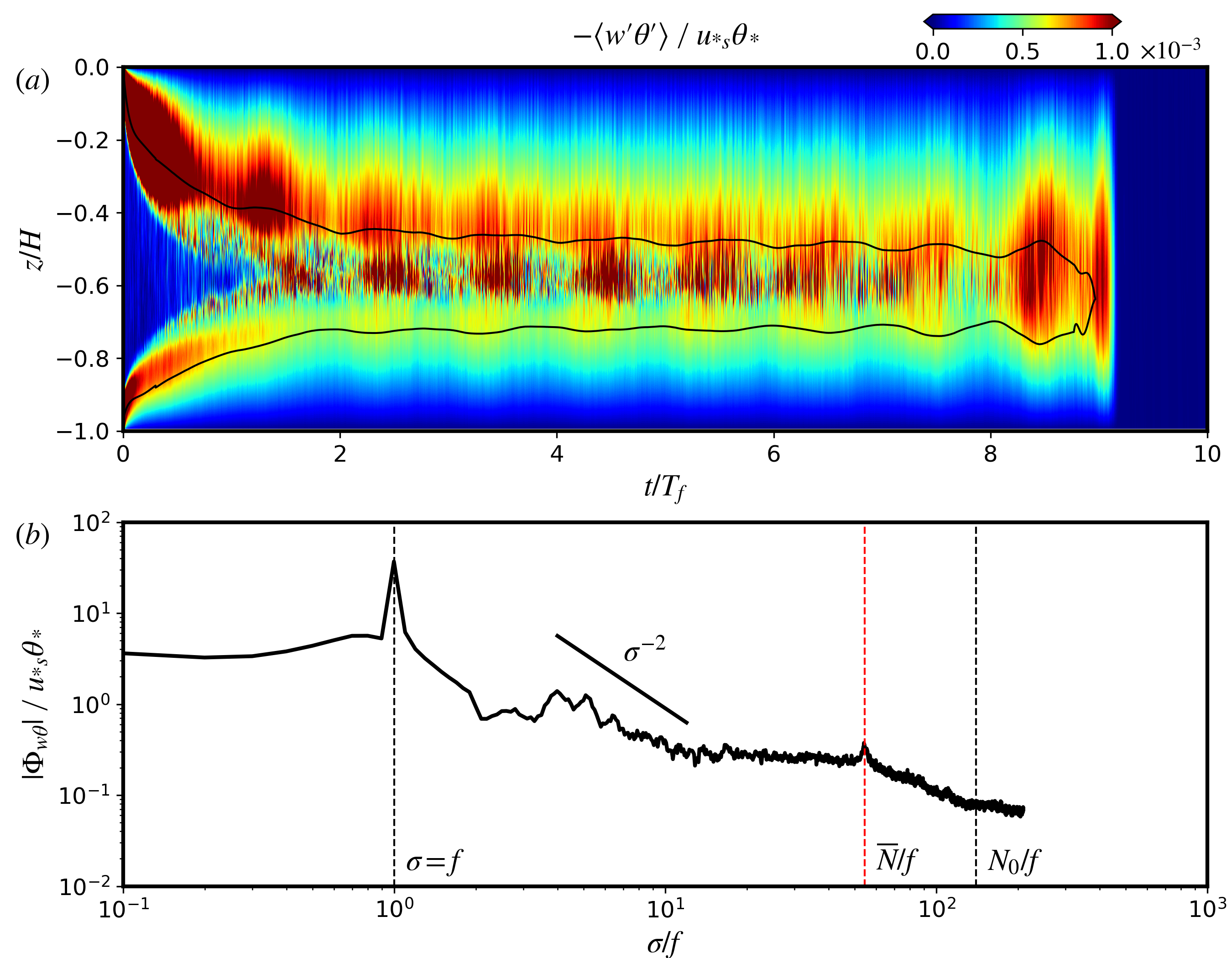}}
    \caption{(\textit{a}) Hovm\"{o}ller diagram of the {\color{red}resolved} buoyancy flux $-\langle w^{\prime} \theta^{\prime} \rangle$ and (\textit{b}) the amplitude of the cross-spectral density $|\Phi_{w\theta}|$ at $z/H=0.6$ as a function of frequency $\sigma$. The black solid lines in (\textit{a}) mark the surface and bottom mixed layer depths. The black dashed lines represent the inertial frequency $f$ and initial buoyancy frequency $N_0$, respectively, and the red dashed line represents the mean buoyancy frequency $\overline{N}$ over the last inertial period before the merger (i.e. $8 < t/T_f < 9.2$). The slope line indicates a $\sigma^{-2}$ rate described by the Garrett-Muck spectrum.}\label{fig:wT}
\end{figure*}

We can examine the contribution of internal waves to energy transfer by looking at the vertical buoyancy flux $-\langle w^{\prime} \theta^{\prime} \rangle$ (as we assume a linear relationship between potential temperature $\theta$ and water density $\rho$). We notice that the strong temperature inversion in the stratified layer will induce notable SGS buoyancy flux there, but the magnitude of the SGS component becomes increasingly small in the final periods before the merger (not shown) and the resolved component captures a large portion (if not the vast majority) of the heat transport. Figure \ref{fig:wT}\textit{a} shows the Hovm\"{o}ller diagram of the resolved buoyancy flux $-\langle w^{\prime} \theta^{\prime} \rangle$, normalized by $u_{*s}\theta_*$. The buoyancy flux disappears after $t/T_f=9.2$, suggesting that the flow turns into neutral condition after that moment. Within the stratified region (the region between the two black thin lines), we can clearly observe two types of fluctuations, i.e. long-time variations and fast-time fluctuations. The long-time variations in the stratified layer have a period of $t/T_f=1$, suggesting that the buoyancy flux is strongly modulated by inertial oscillations. The fast-time fluctuations are associated with the internal gravity waves. It should be noted that a nonzero $-\langle w^{\prime} \theta^{\prime} \rangle$ is not generally expected for oceanic internal waves, and where it occurs it is often associated with internal wave breaking. In case $\mathrm{USF_{S\&B}}$, local density overturns can be seen within the thermoclines (not shown). The vertical mixing caused by these overturns may contribute to the non-zero heat flux in the stratified layer.

Over the last inertial cycle before the merger ($8<t/T_f<9$), even though the corresponding temperature difference is very small, we notice an enhanced heat transfer in the vertical column that ultimately eliminated the internal stratification that leads to the merger. Right after the merger ($9<t/T_f<9.2$), there is another sudden burst in heat flux before its final shutdown. This is probably caused by the enhanced turbulence working to eliminate the temperature difference between the upper and lower regions of the newly formed merged boundary layer.

To quantify the frequency distribution, we transform the time history of $w^{\prime}$ and $\theta^{\prime}$ ($t/T_f<10$) at different vertical levels into frequency space, and introduce the cross-spectral density of $w^{\prime}$ and $\theta^{\prime}$ (shown in figure \ref{fig:wT}\textit{b}) defined as, 
%\begin{equation} \label{eq:wT_spectrum}
%   \langle \widehat{ w^{\prime} \theta^{\prime}} \rangle  = \langle \widehat{w}^{\prime} (x,y,z,\sigma) \widehat{\theta}^{\prime *} (x,y,z,\sigma) \rangle
%\end{equation}
\begin{equation} \label{eq:wT_spectrum}
    \Phi_{w\theta}(z, \sigma) = \langle \widehat{w}^{\prime} (x,y,z,\sigma) \widehat{\theta}^{\prime *} (x,y,z,\sigma) \rangle
\end{equation}
About 4200 samples at each vertical level of $w$, and $\theta$ are collected. To increase the statistical samples, the amplitude of the cross-spectral density $|\Phi_{w\theta}|$ have been averaged over the horizontal plane indicated by the angled brackets, and * denotes the complex conjugate. The frequency is normalized by the inertial frequency $f$. The spectra at $z/H=-0.6$ (figure \ref{fig:wT}b) has a peak value centered at the inertial frequency as expected. The drop in energy levels at intermediate frequencies ($3 < \sigma/f < 15$) approximately follows the Garrett-Muck spectrum with a slope of -2 \citep{Garrett1972GFD}. The spectrum at high frequencies is characterized by an almost horizontal plateau, followed by a drop in power. The energy spectral amplitude has a local bump at $\sigma/f \approx 55$, which coincides with the mean buoyancy frequency $\overline{N}/f$ in the stratified layer (at $z/H=-0.6$) over the last inertial cycle before the merger (i.e. $8 < t/T_f < 9.2$, red dashed line). This peak is related to the final stratification in the last inertial period before the merger. Since the frequency of internal waves cannot exceed the buoyancy frequency $N$, any faster fluctuations at frequencies above $N$ would be purely turbulence, whose energy is quickly dissipated (already filtered here). Above all, the fast-time fluctuations in figure \ref{fig:wT}\textit{a} are due to internal gravity waves and turbulence generated by them.

\begin{figure*}
    \centerline{\includegraphics[width=1.0\linewidth]{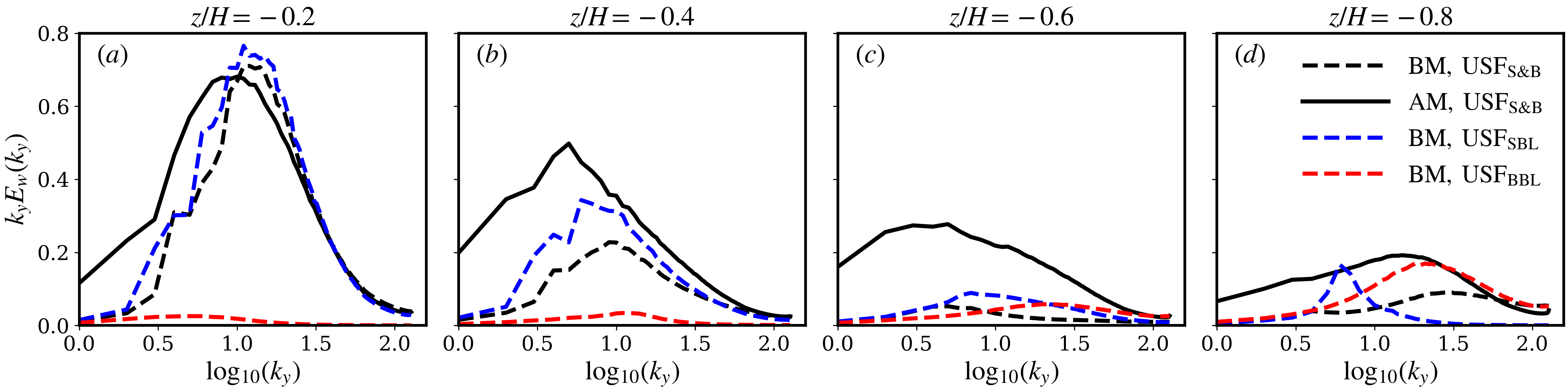}}
    \caption{Wavenumber spectra of the vertical velocity $w/u_*$ at 4 different vertical levels (i.e. $z/H=-0.2,\ -0.4,\ -0.6,\ -0.8$) for all simulations, see (\textit{d}) for legend.}\label{fig:waveSpec}
\end{figure*}

\subsection{Wavenumber spectra analysis} \label{sec:spectra}

The imposition of surface-forcing and geostrophic current significantly alters the turbulent dynamics and spectral cascade. To assess the spectral distribution of energy, figure \ref{fig:waveSpec} shows the one-dimensional wavenumber spectra (calculated in the crosswise $y-$direction and pre-multiplied by the crosswise wavenumber $k_y$) for the vertical velocity $w$ at four vertical levels $z/H=-0.2,\ -0.4,\ -0.6,\ -0.8$ for all simulations. The spectral amplitudes have been averaged over the specified inertial period. As expected, the spectral energy peaks at larger scales in the surface mixed layer due to larger-scale Langmuir structures (figure \ref{fig:waveSpec}\textit{a}), while small-scale stuctures are the most energetic part of turbulence in the bottom mixed layer (figure \ref{fig:waveSpec}\textit{d}). Within the mixed layer, the isolated boundary layer cases ($\mathrm{USF_{SBL}}$ and $\mathrm{USF_{BBL}}$) have larger vertical velocity variance (VVV) over a wide range of scales as compared to the pre-merger state of $\mathrm{USF_{S\&B}}$. This is because more energy are transferred to the potential energy in simulation $\mathrm{USF_{S\&B}}$ due to stronger stratification. Also, modifications of the energy spectra imply the coupling between surface and bottom dynamics. VVV is significantly enhanced especially at larger scales at all depths after the merger for simulation $\mathrm{USF_{S\&B}}$ (black solid lines in figure \ref{fig:waveSpec}), consistent with the vertical distribution of $\overline{\langle w^{\prime}w^{\prime} \rangle}$ in figure \ref{fig:mean_variance}\textit{c}. Additionally, the spectra are left-shifted to the low wavenumber end after merger in the verical column, which is attributed to the effect of full-depth Langmuir circulations. $\mathrm{USF_{SBL}}$ and $\mathrm{USF_{BBL}}$ also yield somewhat VVV at larger scales outside the mixed layers (red line in figure \ref{fig:waveSpec}\textit{a} and \textit{b}, and blue line in figure \ref{fig:waveSpec}\textit{d}), which are likely associated with radiated internal waves. The spectral energy within the transition layer of $\mathrm{USF_{S\&B}}$ (black dashed line in \ref{fig:waveSpec}\textit{c}) also indicate the presence of internal waves, superimposed by small-scale turbulence (at the high-wavenumber end of the spectrum). Consistent with that describe in section \ref{sec:turb}\ref{sec:budget}, $\mathrm{USF_{SBL}}$ prompts energy carried by internal waves (in the stratified layer) five times greater than $\mathrm{USF_{BBL}}$ does, suggesting the SBL turbulence plays a more important role in the generation of internal waves in $\mathrm{USF_{S\&B}}$. However, the scenario would be different depending on the relative magnitude of the surface and current forcing conditions, which is out of the scope in this study.

% --
\section{Conclusions} \label{sec:conclusion}

In this study, we have explored the boundary layer evolution and turbulent structures in an intermediate-depth ocean by means of LES. The wave-averaged equations, with the inclusion of planetary rotation and buoyancy effects, are solved numerically inside a periodic domain of constant water depth, in which the uniformly stratified fluid is driven by a surface forcing (i.e. a constant wind stress and a monochromatic surface wave) and a steady current in geostrophic balance. The latter is generated by an imposed pressure gradient applied in the crosswise direction. We refer to the resulting flow as overlapping boundary layers since the SBL and BBL co-exist.

Over the course of development, the overlapping boundary layers evolve through 3 phases separated by two transitions. In phase 1 ($t/T_f<1.5$), the co-existing boundary layers grow by entrainment at the same rate as their isolated counterparts. The water temperature exhibits a five-layer structure. Two pycnoclines, which form at the edges of the upper and lower mixed layers, are separated by an interior stratified region. The interior stratification inhibits the vertical turbulent exchange, but it also provides a necessary condition for the generation of internal waves. The SBL and BBL partly communicate by virtue of vertically propagating internal waves. Transition 1 occurs when the two pycnoclines merge into one, and the stratification significantly increases by a factor of 5. 

In phase 2, the boundary layer growth is stalled and the flow field is delimited by 3 distinct regions in the vertical column. These regions include the surface mixed layer where Langmuir turbulence dominates, the stratified layer where turbulence is energized by energy flux carried by internal waves, and the bottom mixed layer where bottom-generated turbulence dominates. In our case, the internal waves are mainly excited by Langmuir cells in the SBL, and they modulate turbulence in the BBL (based on conditionally averaged results and TKE budget), so that the energy transfer is from top to bottom (but this could possibly be different depending on the strength of surface and current forcings). In this phase, the interior stratification is slowly eroded by downward heat fluxes that cool the SBL and warm the BBL. Coriolis seems to play a critical role in this phase, as the heat fluxes are strongly modulated by inertial oscillations. Transition 2 occurs when the interior stratification is completely eroded by vertical mixing, causing the two boundary layers to finally collapse into one.

In phase 3, as the two boundary layers are fully merged, Langmuir circulations are found to extend down to the bottom wall, even though the water is quite deep in our case (Stokes drift vanishes in the lower half of the vertical column). The full-depth Langmuir circulations promote the vertical momentum transfer and enhance the bottom shear stress, leading to increased contribution from the shear production (near the bottom) and Stokes production (near the surface), which are the main causes for the drastic enhancement of turbulence levels after the merger. From the TKE budget analysis, the energy is transferred from the surface part to the bottom part via non-local transport possibly due to the full-depth Langmuir circulations, but pressure transport redistribute the energy in the vertical.

In this study, our major intent is to characterize the boundary layer development in an finite-depth ocean and to quantify how the SBL and the BBL interact with each other. We are aware that only a limited set of typical ocean conditions are considered here, while a full understanding of how Langmuir turbulence interacts with the bottom shear under varying wind-wave-current forcing conditions (e.g. oblique forcing) warrants further investigations. 

%vs

% \section{Section title}
% \subsection{subsection one}
% text...
% \subsection{subsection two}
% \section{Section title}

%%%
% \section{First primary heading}

% \subsection{First secondary heading}

% \subsubsection{First tertiary heading}

% \paragraph{First quaternary heading}

%%%%%%%%%%%%%%%%%%%%%%%%%%%%%%%%%%%%%%%%%%%%%%%%%%%%%%%%%%%%%%%%%%%%%
% ACKNOWLEDGMENTS
%%%%%%%%%%%%%%%%%%%%%%%%%%%%%%%%%%%%%%%%%%%%%%%%%%%%%%%%%%%%%%%%%%%%%
\acknowledgments
This work is supported by the ARPA-E MARINER Program (DE-AR0000920).

%%%%%%%%%%%%%%%%%%%%%%%%%%%%%%%%%%%%%%%%%%%%%%%%%%%%%%%%%%%%%%%%%%%%%
% DATA AVAILABILITY STATEMENT
%%%%%%%%%%%%%%%%%%%%%%%%%%%%%%%%%%%%%%%%%%%%%%%%%%%%%%%%%%%%%%%%%%%%%
% 
%
\datastatement
The LES code and relevant materials that support the findings of this study can be obtained from https://github.com/GAbelois/OverlappingBLs.git.

%%%%%%%%%%%%%%%%%%%%%%%%%%%%%%%%%%%%%%%%%%%%%%%%%%%%%%%%%%%%%%%%%%%%%
% APPENDIXES
%%%%%%%%%%%%%%%%%%%%%%%%%%%%%%%%%%%%%%%%%%%%%%%%%%%%%%%%%%%%%%%%%%%%%
%
% Use \appendix if there is only one appendix.
%\appendix

% Use \appendix[A], \appendix[B], if you have multiple appendixes.
%\appendix[A]

%% Appendix title is necessary! For appendix title:
%\appendixtitle{}

%%% Appendix section numbering (note, skip \section and begin with \subsection)
% \subsection{First primary heading}

% \subsubsection{First secondary heading}

% \paragraph{First tertiary heading}

%% Important!
%\appendcaption{<appendix letter and number>}{<caption>} 
%must be used for figures and tables in appendixes, e.g.,
%
%\begin{figure}
%\centerline\includegraphics[width=19pc,angle=0]{figure01.pdf}\\
%\appendcaption{A1}{Caption here.}
%\end{figure}
%
% All appendix figures/tables should be placed in order AFTER the main figures/tables, i.e., tables, appendix tables, figures, appendix figures.
%
%%%%%%%%%%%%%%%%%%%%%%%%%%%%%%%%%%%%%%%%%%%%%%%%%%%%%%%%%%%%%%%%%%%%%
% REFERENCES
%%%%%%%%%%%%%%%%%%%%%%%%%%%%%%%%%%%%%%%%%%%%%%%%%%%%%%%%%%%%%%%%%%%%%
% Make your BibTeX bibliography by using these commands:
\bibliographystyle{ametsoc2014}
\bibliography{references}

%%%%%%%%%%%%%%%%%%%%%%%%%%%%%%%%%%%%%%%%%%%%%%%%%%%%%%%%%%%%%%%%%%%%%
% TABLES
%%%%%%%%%%%%%%%%%%%%%%%%%%%%%%%%%%%%%%%%%%%%%%%%%%%%%%%%%%%%%%%%%%%%%
%% Enter tables at the end of the document, before figures.
%%
%

%%%%%%%%%%%%%%%%%%%%%%%%%%%%%%%%%%%%%%%%%%%%%%%%%%%%%%%%%%%%%%%%%%%%%
% FIGURES
%%%%%%%%%%%%%%%%%%%%%%%%%%%%%%%%%%%%%%%%%%%%%%%%%%%%%%%%%%%%%%%%%%%%%
%% Enter figures at the end of the document, after tables.
%%
%

\end{document}